\documentclass[aps,prd,twocolumn,superscriptaddress]{revtex4-2}
\usepackage[english]{babel}
\usepackage{url}
\usepackage{balance}
\usepackage{lineno}

\usepackage{amsmath,amssymb}
\usepackage{graphicx}
\usepackage{cancel}
\usepackage{comment}
\usepackage[normalem]{ulem}
\usepackage[colorlinks=true, allcolors=blue]{hyperref}
\usepackage[dvipsnames]{xcolor}

\begin{document}

 \author{F.~Arias-Arag\'on}
 \affiliation{Istituto Nazionale di Fisica Nucleare, Laboratori Nazionali di Frascati, Frascati, 00044, Italy}
 \author{L.~Darmé}
 \affiliation{Universit\'e Claude Bernard Lyon 1, CNRS/IN2P3, Institut de Physique des 2 Infinis de Lyon, UMR 5822, F-69622, Villeurbanne, France}
 \author{R.~Gargiulo}
 \affiliation{Sapienza University, Department of Physics, Rome, 00185, Italy}
 \affiliation{INFN, Sezione di Roma, Rome, 00185, Italy}
 \author{G.~Grilli~di~Cortona}\affiliation{Istituto Nazionale di Fisica Nucleare, Laboratori Nazionali del Gran Sasso, Assergi, 67100, L'Aquila (AQ), Italy}
\author{V. Kozhuharov}
\affiliation{Istituto Nazionale di Fisica Nucleare, Laboratori Nazionali di Frascati, Frascati, 00044, Italy}
\affiliation{Faculty of Physics, Sofia University “St. Kl. Ohridski”, 5 J. Bourchier Blvd., 1164 Sofia}
\author{E.~Nardi}
\affiliation{Istituto Nazionale di Fisica Nucleare, Laboratori Nazionali di Frascati, Frascati, 00044, Italy}
\affiliation{Laboratory of High Energy and Computational Physic, HEPC-NICPB, R\"avala 10, 10143, Tallin, Estonia}
\author{M. Raggi}
\thanks{Corresponding author}
\email{mauro.raggi@cern.ch}
\affiliation{Sapienza University, Department of Physics, Rome, 00185, Italy}
\affiliation{INFN, Sezione di Roma, Rome, 00185, Italy}
\affiliation{CERN, Geneva, Switzerland}

\author{T.~Spadaro}
\affiliation{Istituto Nazionale di Fisica Nucleare, Laboratori Nazionali di Frascati, Frascati, 00044, Italy}
\author{P.~Valente}
 \affiliation{INFN, Sezione di Roma, Rome, 00185, Italy}

 \title{Dark sector searches with high-intensity positron beams in the CERN North Area}

\begin{abstract}
Dark sector models present a rich phenomenology that requires high-intensity beams and precision detectors for thorough exploration. 
The NA62 experiment has already published several constraints on dark sector models, leveraging proton beam dump and meson decay techniques. This proposal aims to demonstrate the NA62 detector discovery potential for dark sector candidates, by using the positron-on-target technique. High-intensity secondary positron beams, reaching up to $\sim$150 GeV energy, have already been produced at the North Area extracted beam lines. If a positron beam with an intensity in the range of 2$\times10^{14}$ positrons on target per year is delivered, the NA62 detector would be ideal for  searches of dark sector particles in both visible and invisible decay channels. Additionally, positron on target collisions would enable precision measurements of key standard model observables, including a detailed scan of $\sigma(e^+e^-\to\pi^+\pi^-$) and $\sigma(e^+e^-\to\mu^+\mu^-)$ at the di-pion and di-muon production threshold, with discovery potential for the True Muonium ($\mu^+\mu^-$) bound state. 
\end{abstract}
\maketitle
\newpage

{
\hypersetup{linkcolor=black}
\setcounter{tocdepth}{2}
}

\newpage
\section{Introduction}
Accelerating primary positrons to the energy range of hundreds of GeV is a challenging and costly endeavor.
In 2018, an Expression Of Interest (SPSC-EOI-018) was submitted to the CERN SPSC Scientific Committee for the development of an electron beam facility at the Super Proton Synchrotron (SPS) accelerator. 
The project proposed generating an extracted high-intensity 16 GeV electron beam, after accelerating electrons in the SPS~\cite{Akesson:2640784}.
In contrast, secondary electron or positron beams, produced as by-products of proton collisions, can benefit from the existing high-energy proton machines at CERN. 
At CERN’s North Area (NA), extracted beam lines routinely produce electrons with energies $\approx$100 GeV from the 400 GeV protons of the SPS for the NA64 experiment~\cite{NA64:2023wbi}. However, the intensity and purity of the secondary positron beams have so far been the main limiting factors for positron-on-target experiments~\cite{Andreev:2023xmj}.
 
Positron-on-target collisions have proven to be an effective strategy for dark sector particles searches, thanks to the increased number of available production mechanisms, in particular the resonant annihilation of positron-electron into a dark sector boson~\cite{Nardi:2018cxi}. 
More recently, it has been highlighted that the motion of electrons within high-$Z$ target materials allows experiments to significantly surpass the center-of-mass (c.m.) energy limit 
 $\sqrt{2 m_e E_{\text{Beam}}}$~\cite{Arias-Aragon:2024qji,Arias-Aragon:2024gpm} of annihilation-driven processes, which is typically imposed by the electron-at-rest condition in fixed-target experiments. 
 In addition, the ability of associated production to enhance dark sector particle production rates in intermediate mass ranges makes positrons the preferred beam over electrons.
Finally, to explore lepto-philic dark sector models,  lepton-based production mechanisms are required, making searches with positron beams complementary to those based on protons. 

In this paper, we will illustrate  
how high-energy and high-intensity positron beams can be produced in the CERN North Area, and how a new project NA62e+, reusing the NA62 detector, can achieve outstanding sensitivity to dark sector particles across a variety of scenarios.  

All limits presented in this proposal are evaluated under the background-free hypothesis. More detailed simulations will be performed in future studies, once the feasibility and the maximum achievable intensity of the beam line have been clarified.

The exclusion limits obtained will complement those of the BDF-SHiP program \cite{SHiP:2015vad}, both in terms of model constraints and the coverage of the energy–coupling parameter space.

Using a positron beam with a momentum of 80–100~GeV, NA62e+ will also be able to measure the hadronic cross section in the $\sqrt{s}$ range up to 300~MeV, where measurements at circular colliders suffer from particularly large statistical and systematic uncertainties.
These cross sections, particularly $\sigma (e^+e^- \to \pi^+\pi^-)$, are well-known to be crucial theoretical inputs for the prediction of  $(g-2)_\mu$.

By adjusting the positron beam to a slightly lower momentum of $\sim$45 GeV, the NA62e+ experiment will be able to make precise measurements of the cross section for $e^+e^- \to \mu^+\mu^-$, which has been proposed as a muon production scheme for the muon collider~\cite{Antonelli:2015nla}.
Additionally, during a dedicated run using a multi-target system and an upstream tracker, NA62e+ could potentially achieve the first observation of true muonium atoms, using techniques similar to the ones in~\cite{h4}. 

\section{The North Area beam lines and the NA62 detector}
In this section, we will describe the key aspects of the North Area beam line and NA62 detectors that are relevant to the following discussion. 

\begin{figure}[h]
 \begin{center}
 \resizebox{0.9\linewidth}{!}{%
   \includegraphics{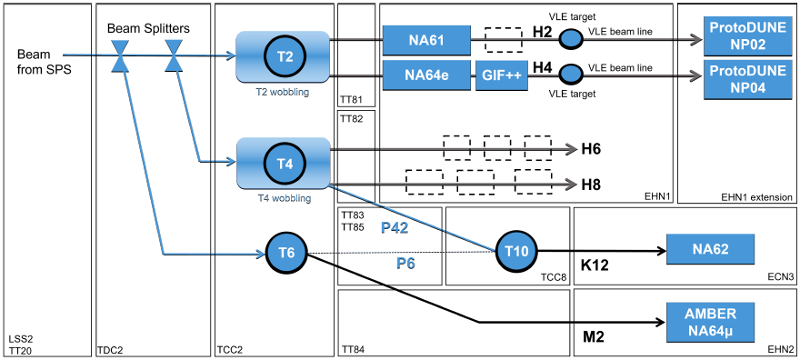}
 }
 \caption{The schematic of the CERN North area extracted beam-lines~\cite{Ahdida:2023okr}. 
 }
 \label{fig:NorthArea}    
 \end{center}
 \end{figure}

The ECN3 experimental hall, located 15 m underground (see Fig.~\ref{fig:NorthArea}), is part of the North Area complex, at the CERN Prevessin site. The complex is composed of a number of experimental facilities and lines making use of secondary and tertiary beams produced from the high-intensity proton beam extracted from the SPS; in particular, ECN3 is served by the K12 beamline.
After the slow extraction from SPS, protons are split into three branches by means of two beam splitters, directed towards three different targets: T2, T4, and T6, located in the TCC2 target hall (see Fig. \ref{fig:NorthArea}).
Primary protons not interacting in the T4 target are transported by the P42 transfer line over almost 900 metres to the T10 target located in the target hall TCC8. The T10 target is the starting point of the K12 beam line which delivers the high-intensity, unseparated secondary hadron beam to the NA62 experiment, located in the downstream experimental hall ECN3.

A schematic of the CERN North Area beam lines is shown in Fig.~\ref{fig:NorthArea}.
A more comprehensive description of the P42 and K12 beam lines can be found in~\cite{Banerjee:2774716}. The T6 target system and the M2 beam line produce the high-intensity muon beam for the NA64$\mu$, MUonE, and AMBER experiments.
The T6 target receives the highest protons flux among all the North Area targets, with up to $\sim1.5\times 10^{13}$ protons per pulse.

\subsection{The K12 beam line}
\begin{figure}[h]
 \begin{center}
 \resizebox{0.9\linewidth}{!}{%
   \includegraphics{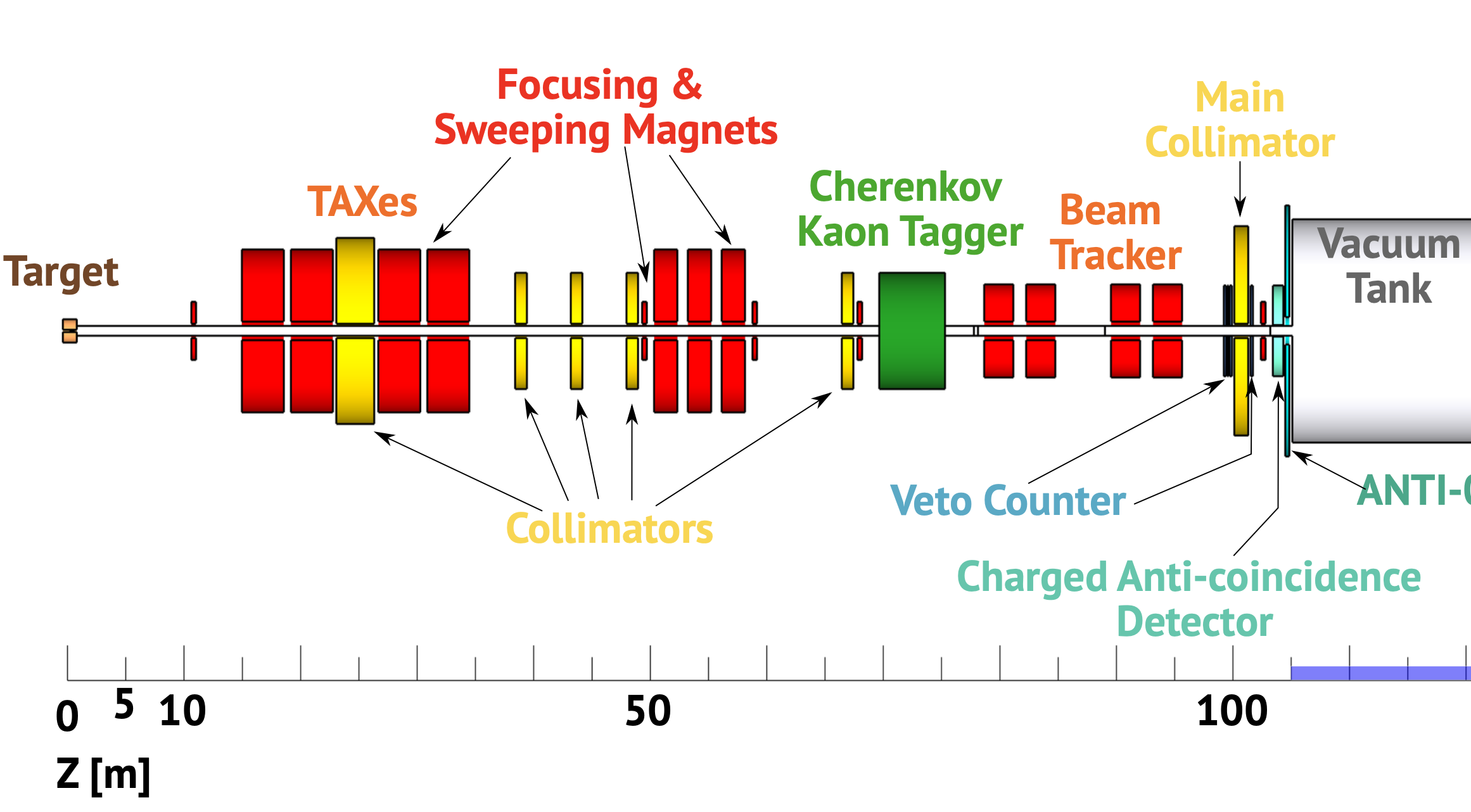}
 }
 \caption{A schematic view of the K12 beam-line~\cite{HIKE:2023ext}}
 \label{fig:K12}    
 \end{center}
 \end{figure}

The K12 beam line is located downstream of the P42 transfer line. 
It has been designed to provide a high intensity 75 GeV kaon beam to the NA62 experiment.~Fig. \ref{fig:K12} represents a schematic view of the K12 beam line.
The K12 mixed hadron beam is produced by the interaction of the SPS primary protons with the T10 target made of 4$\times$100~mm long beryllium rods.

The beam is then focused onto a pair of dump collimators known as TAX (``Target Attenuator Experimental areas") made of massive copper and steel blocks. A set of four strong dipoles surrounding the TAX (the ``first achromat") ensures the selection of secondary particles with a momentum of 75~GeV with a 1.1\% RMS resolution. Off-momentum and neutral particles are absorbed into the TAX. Positrons produced in the proton collisions are attenuated using a thin tungsten converter.
After additional sweeping magnets and collimators are used to reduce muon halo, the kaon component of the unseparated hadron beam is identified by the KTAG Cherenkov detector.
In the ``second achromat" region, a section of the beam is deflected downward by a set of dipoles, and its momentum is measured by four silicon pixel detectors known as the GigaTracker with a resolution of 0.2\%. A thorough description of the beam line can be found in~\cite{Banerjee:2774716}. 

\subsection{The NA62 experiment}

\begin{figure}[h]
 \begin{center}
 \resizebox{1.\linewidth}{!}{%
   \includegraphics{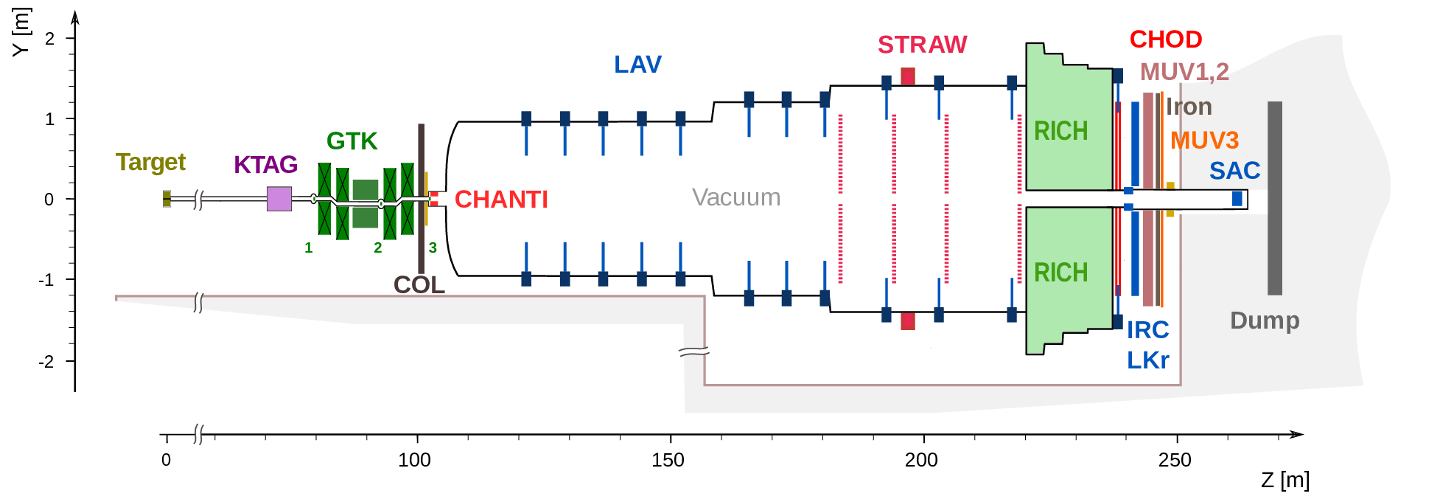}
 }
 \caption{The NA62 experiment layout~\cite{NA62:2017rwk}.}
 \label{fig:NA62}    
 \end{center}
 \end{figure}

The NA62 experiment, located in CERN's North Area, is a fixed-target experiment designed to measure the extremely rare decay $K^+ \to \pi^+ \nu \bar{\nu}$. With a SM branching ratio as small as $8 \times 10^{-11}$, this decay requires a background rejection of the order of $\sim 10^{11}$ to be identified. To achieve this challenging goal, NA62 is equipped with high-precision, redundant detectors, which can be categorized into several groups:

\begin{itemize}
    \item Particle identification detectors: KTag, RICH, LKr calorimeter 
    \item Photon veto detectors: LAV, LKr and SAC, IRC 
    \item Charged particle veto detectors: CHANTI, MUV, HASC
    \item Energy and momentum measuring detectors: LKr calorimeter and straw tracker
\end{itemize}
The layout of the detector is shown in Fig.~\ref{fig:NA62}. Kaon decays are reconstructed within a vacuum decay region over 70 meters long. The detector boasts an unprecedented photon and charged track rejection capability, below the $1 \times 10^{-4}$ level, and enables precise measurement of the 4-momentum of both photons and charged particles. The particle identification system, consisting of the RICH detector, downstream calorimeters, and muon veto, allows for pion-muon-electron separation down to the $10^{-5}$ level.
 
 The NA62 detector has already been successfully employed for dark sector searches, both in kaon decays~\cite{NA62:2020xlg,NA62:2020pwi} and in dump mode~\cite{NA62:2023nhs}. The current DAQ system can handle a Level~0 trigger rate of over 10~MHz, which is expected to be significantly higher with respect to positron mode operation requirements. A comprehensive description of the detector and its performance is beyond the scope of this paper; however, detailed information on individual detector components and their performance can be found in~\cite{NA62:2017rwk}.

\section{Positron beams in the CERN North area}

This section aims to provide a rough estimate of the number of positrons that could be produced in the CERN North Area (NA) during the Beam Dump Facility (BDF/SHiP) era after the LHC Long Shutdown 3 (LS3). 
Positron beams are already routinely produced e.g. from the T2 target and have been successfully used by the NA64 experiment~\cite{NA64:2023ehh} in the H4 beam line (see Fig.~\ref{fig:NorthArea}).
Nevertheless, the number of protons per pulse is constrained by radiation protection requirements in the surface hall and by the need to share the T2 target among two beam lines, which also limits the maximum deliverable flux relative to the potential production capacity.
The availability of positron beams motivates the study of the physics goals of a positron-on-target experiment that could fully exploit the maximum available production rate, assuming a dedicated beamline with a proton rate of a few~$\times 10^{13}$ protons per pulse (ppp).
Identifying production techniques and optimizing hadron contamination will require a dedicated study by the CERN accelerator division and is outside the scope of this paper. The figures presented here are intended only as indicative estimates of the positron flux for the subsequent physics case study.

\subsection{Positron production rates in targets}

To estimate the number of positrons that could be produced in the CERN NA, we start with the annual proton delivery figures outlined in the recent report ``\emph{Post-LS3 Experimental Options in ECN3}"~\cite{Ahdida:2023okr}. The last 3 rows of Tab.~\ref{tab:NProtons} summarize the data from Table 1 of~\cite{Ahdida:2023okr}. 
In the first row, we included an estimate of the current maximum achievable proton flux, based on the flux reached at T10 during the NA62 proton beam dump operation in 2024 and the total number of spills delivered during the 2024 run (approximately $600\times 10^3$).

\begin{table}[htpb!]
\centering
\begin{tabular}{l|cc|c}
\hline
 & Intensity on T10  &  $N_p$/year & N$e^+$/y    \\ 
 &[$10^{13}$ ppp] &[$10^{19}$] & in 1$\mu$sr [$\times 10^{13}$]\\
\hline
NA62 ($K^+$) & 0.55 &0.33 & 1.2\\
\hline
HIKE ($K^+$) & 1.  & 0.72 & 2.5 \\ 
HIKE ($K^0$) & 1.2 & 1.2 & 4.2 \\ 
BDF & 4.0 & 4.0  & 14\\ 
\hline
\end{tabular}
\caption{Present and expected proton beam intensities (ppp) and positrons production rates per $\mu$ster. Positron rates are obtained using 500~mm Be target thickness~\cite{Ahdida:2023okr} and $600\times 10^3$.}
\label{tab:NProtons}
\end{table}

\begin{figure}[h]
 \begin{center}
 \resizebox{0.8\linewidth}{!}{%
   \includegraphics{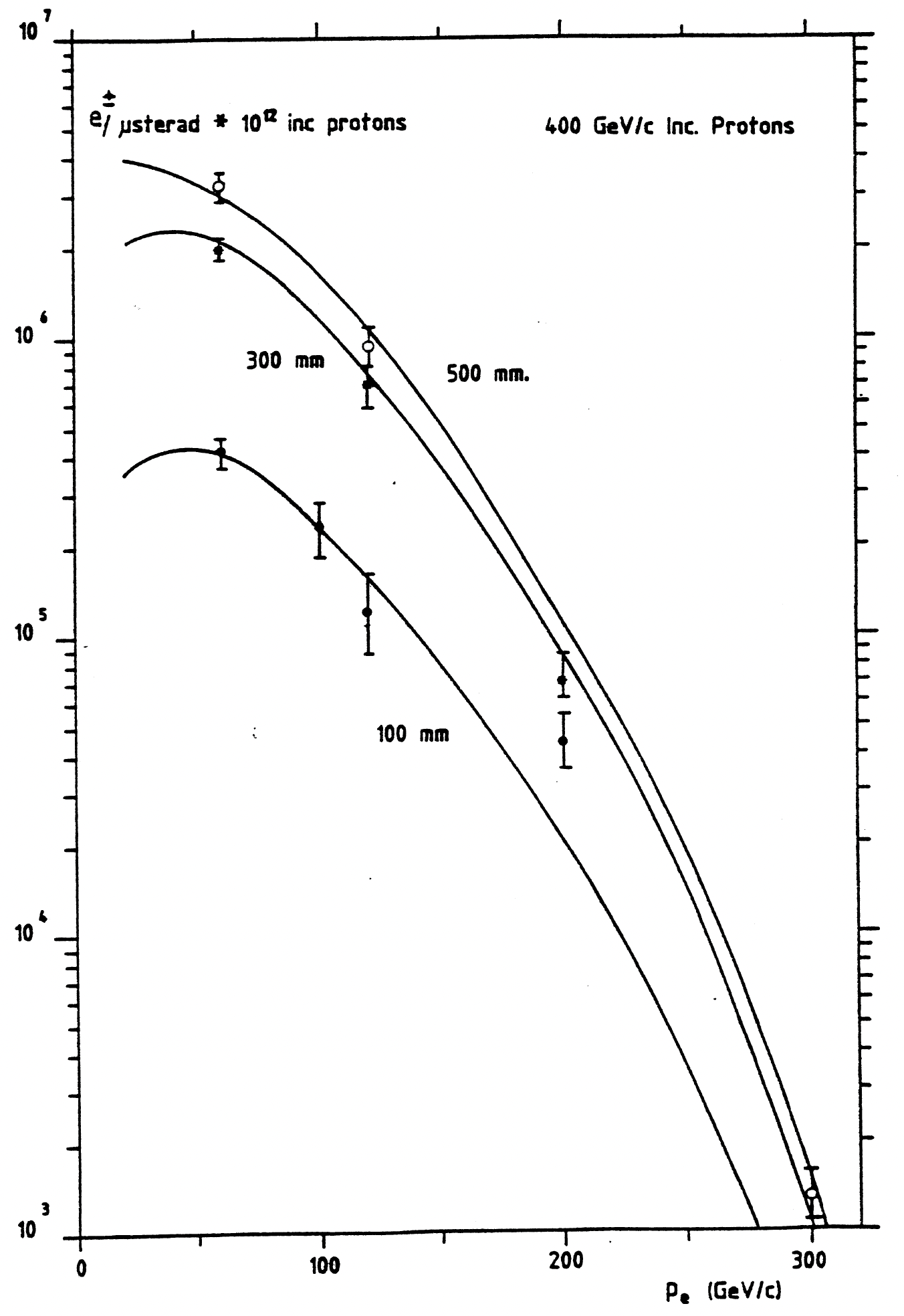}
 }
 \caption{Absolute electron/positron production rates in 1 $\mu$sr per 1$\times 10^{12}$ 400~GeV protons from Beryllium target~\cite{Doble:164934}. Lines represents simulations while open dots represent experimental measurements with 500 mm Be target.}
 \label{fig:PosProd}    
 \end{center}
 \end{figure}

The next ingredient for our calculation is the positron yield per proton-on-target. 
Fig.~\ref{fig:PosProd} shows the positron rate obtained using the TURTLE simulation program, compared to experimentally measured values at CERN~\cite{Doble:164934} for a $500$ mm beryllium target, very similar to those currently used in the T4, T6 and T10 targets.
MC predictions are in good agreement with the measured values.

At the nominal NA62 beam energy of 75 GeV, the yield would be similar to the open dot at 60~GeV in Fig.~\ref{fig:PosProd} approximately $3.5 \times 10^6$ $e^{\pm}/\mu$sr per $1\times10^{12}$ protons on a 500~mm Be target.  This corresponds to a positron production probability per proton per $\mu$sr of $e^+/p = 3.5 \times 10^{-6}$. Using this probability, we calculate the number of positron produced per year (N$_{e^+}$/y) within a $\mu$sr acceptance, as reported in the last column of Tab.~\ref{tab:NProtons}. 

\subsection{Un-separated and separated positron beams} 

\begin{figure}[h]
 \begin{center}
 \resizebox{0.8\linewidth}{!}{%
   \includegraphics{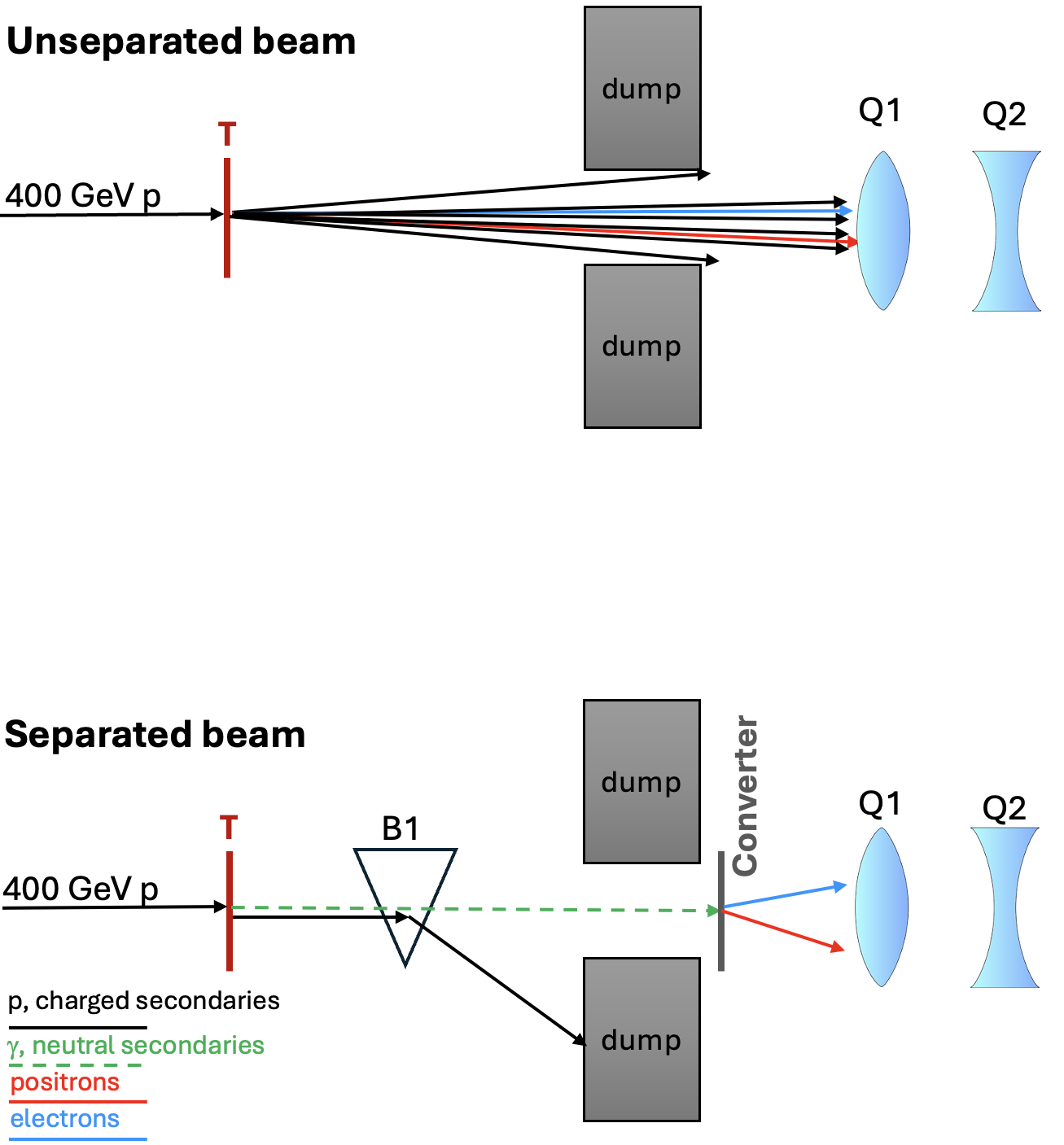}
 }
 \caption{Left: production scheme of unseparated and separated positron beams.}
 \label{fig:BeamTypes}    
 \end{center}
 \end{figure}

Since the late `70s, two strategies were identified for producing positrons out of the SPS protons in the CERN North Area.
The layout of generic beams obtained in the two configuration is shown in Fig.~\ref{fig:BeamTypes}.

The simplest method, which produces the so-called ``unseparated" beam, relies on the positron yield generated by proton on target collisions. The dominant mechanism is the production of $\pi^0$ mesons, which promptly decay into two photons. These photons then convert into $e^+e^-$ pairs either within the target material itself or in a dedicated converter.
The material type and thickness for the production target must be optimized to maximize the positrons-to-hadrons ratio. 
A key parameter is the ratio of radiation length to nuclear interaction length, while the balance between positron production and absorption determines the optimal converter thickness.
Studies conducted at the NA extracted beam lines in the `80s concluded that a 500~mm Be target offers the best performance~\cite{Doble:164934}.
A properly designed focusing system positioned between the target and the dump aperture can significantly improve the beam acceptance compared to the purely geometrical configuration illustrated in Fig.~\ref{fig:BeamTypes}.

Although unseparated positron beams are easy to produce and offer very high intensities, their positron content is typically limited to a maximum of about 30\% at $\sim50$~GeV, dropping below the percent level at higher momenta. Producing a pure positron beam requires a separation strategy, typically based on exploiting positron synchrotron radiation.
The separation increases with $E^4$ and is only efficient for momenta above 120-150~GeV/c where the electron rate is much lower that at 75~GeV/c. 

A much lower content of hadrons can be obtained by using ``separated beams" from photon conversion. In this scheme, the unseparated beam obtained after the proton-on-target collisions is deflected by a bending magnet B1 towards a dump, to eliminate all the charged secondaries. The photon component of the neutral beam, after crossing the dump hole, is converted into $e^+e^-$ pairs by a photon converter. This solution provides somewhat lower intensity due the efficiency of the conversion target.    
In positron beams from photon conversion the hadron contamination is dominated by the protons from $\Lambda\to p\pi^-$ and pions from $K_S$ decays. According to~\cite{Andreev:2023xmj}, the hadron contamination at 75 GeV positron beam is $\sim$1\% in the H4 beam line. Unfortunately, the value grows rapidly with the beam energy reaching $\sim$100\% for energy above 150 GeV~\cite{Andreev:2023xmj}, limiting the advantage of using positron-on-target technique in the CERN NA64 experiment. 

\subsection{Positron energy range}

As shown in Fig.~\ref{fig:PosProd}, positron production rates in proton-on-target collisions decrease significantly at high energies. However, production rates of nearly 1~MHz/$\mu$sr can still be achieved for energies up to 200 GeV with \(1 \times 10^{13}\) ppp on a 500 mm Be target. 

In contrast, intensity is not a limiting factor at lower beam energies.  
We assume 40 GeV as the lowest achievable beam energy, based on the requirements of the NA64 positron program at the H4 line in the CERN North Area~\cite{NA64:2023ehh}, which features a positron production mechanisms very similar to those proposed in this paper.  
Nevertheless, lower
energies may also be achievable by exploiting a new beamline design.

Designing a positron beamline with the widest possible energy range would be beneficial for this project, as it would allow NA62e+ to explore different regions of center-of-mass energy, thereby covering a broader range of physics cases. For this study, an energy range of 40–150 GeV is ideal. The expected beam energy spread after passing through an achromat system similar to the one existing in K12 beamline can be as low as the percent level\cite{Hahn:1404985}.

\subsection{Required positron beam parameters}

In this section, we summarize our estimates of the key parameters of a potential future NA positron beam. While more precise calculations will be necessary, the estimates presented here are sufficiently accurate for the current discussion of possible physics programs.

To evaluate the positron yield in the NA, it is necessary to define the angular acceptance of a future beamline. Taking inspiration from the existing M2 beamline, we assume that a beamline with an acceptance of up to 5–6 $\mu$sr is achievable by employing a focusing system after a 500 mm Be target. The T6 target, which serves the M2 line, is routinely operated with protons per pulse (ppp) up to \( 15 \times 10^{12} \). Therefore, we assume a similar proton rate could be achieved on a dedicated positron production target.    

In this scenario, the positron line is expected to use approximately 30\% of the available protons in the post-LS3 era, and could be operated concurrently with the BDF. However, the fraction of protons required could be significantly reduced if a positron beamline with an acceptance greater than 5–6~$\mu$sr can be designed.

We note that the existence of the P6 transfer line, connecting T6 to the P42 line in TTC8 (see Fig. \ref{fig:NorthArea}), would have allowed the use of T6 as a production target and the delivery of positrons to NA62 at the required rate. Unfortunately, to the best of our knowledge, the P6 transfer line was dismantled in the early 2020s \cite{Ahdida:2023okr}. Nevertheless, the option of using T6 as the target and rebuilding the P6 transport line should be considered, if it provides a simpler solution than modifying the P42 line.

Table~\ref{tab:NA62eBeamParam} summarizes expected rates of a 75 GeV positron beam.

\begin{table}[htbp!]
    \centering
    \footnotesize
    \begin{tabular}{|l|c|c|c|c|c|}
        \hline
        Beam config &$N_{ppp}\times10^{12}$&Acc& $N_{e+}/(N_p \times \mu sr)$& $N_{e+}$/y \\ 
    \hline
    \hline
    Unsep. focused &15 &6.0~$\mu$sr& 3.5$\times10^{-6}$ &$\sim 2\times10^{14}$\\
    Separated &15 & 1.5~$\mu$sr& 1.5$\times10^{-6}$& $\sim 2\times10^{13}$\\
    \hline
    \end{tabular}
    \caption{Estimated positron beam parameters using 600$\times10^{3}$ spill/y of $15\times10^{12}$ ppp on a 500 mm Be target.} 
    \label{tab:NA62eBeamParam}
\end{table}

The difference in rates between the unseparated focused beam and the separated beam reflects the lower acceptance as  well as the limited efficiency of the photon converter ($\sim$ 50\% or more). 

Table~\ref{tab:NA62eBeamParam} demonstrates that a focused, unseparated beam is the most effective solution for achieving high positron rates in the NA. Assessing the feasibility and precise positron yield of the two solutions or exploring alternative, potentially more effective approaches is beyond the scope of this paper. Such evaluations require dedicated studies and Monte Carlo simulations conducted by the CERN accelerator division.  

Nevertheless, even if the positron flux were significantly lower than the value estimated in Table~\ref{tab:NA62eBeamParam}, it would still offer a unique opportunity for positron-on-target experiments. In this study, we assume that a flux of \(2.0 \times 10^{14}\) 75~GeV positrons on target, with hadron contamination reduced to less than 1\%, can be delivered to an NA62-like detector. 

\section{Dark sectors at NA62e+ with a 75 GeV positron beam}

With a 75 GeV high-intensity positron beam on target, the existing NA62 detector can support multiple dark sector search techniques simultaneously. Moreover, the beam’s luminosity can be adjusted to match the detector’s rate capability by selecting different target materials and thicknesses, optimized for specific dark sector scenarios.
Even when operating in a surface-level experimental hall, target thicknesses can range from a few hundred microns to several tens of centimeters (in dump mode), thanks to the low radiation levels generated by  $\approx 10^{14}$ positrons on target per year (e$^+$OT/$y$) at 75 GeV. 

\subsection{NA62 luminosity with different target materials and thickness}

We estimate the luminosity
for two target scenarios: one using the existing 500 $\mu$m thick silicon NA62 GiGaTracker3 station as the active target, and another using a 500 $\mu$m tungsten target, either placed close to or significantly downstream of the GiGaTracker3 station to optimize detector acceptance.
The 500 $\mu$m thick tungsten target solution, previously proposed by the LDMX experiment~\cite{LDMX:2018cma}, is complementary to the silicon target scenario. Low-Z and high-Z materials can enhance different contributions to the total cross section relative to specific backgrounds. 
The luminosity per year of data taking is computed as follows:
\begin{equation}
\begin{aligned}
\frac{\text{NA62}_{\text{lumi}}}{y} &= \frac{ e^+\text{OT}}{y} \frac{ \mathcal{N}_A ~ Z ~ \rho }{A} D \\
&= 46.7\,\text{pb}^{-1} \left(\frac{e^+OT/y}{2\times 10^{14}}\right)\left(\frac{Z}{74}\right)\left(\frac{183.84\,\text{g/mol}}{A}\right) \\
& \left(\frac{\rho}{19.3\,\text{g/cm}^3}\right)\left(\frac{D}{0.05\,\text{cm}}\right)\,.
\end{aligned}
\label{eqn:NA62Lumi}
\end{equation}
Here, e$^+$OT/$y=2\times10^{14}$ represents the number of positrons on target per year and $D$ the target thickness in cm. This yields approximately 6.7 pb$^{-1}$/year for the silicon target and around 47 pb$^{-1}$/year for the tungsten target.
Much higher positron rate or thickness of the targets will require a revision of the trigger and DAQ system of the NA62 experiment.
Moreover, even in the e$^+$OT/y = 2 $\times 10^{14}$ configuration, the experiment would benefit from an improved small-angle ($< 8,\mathrm{mrad}$) photon veto system. The current solution limits the single-photon rejection efficiency. Improving the time and energy resolution and the rate capability of the IRC and SAC detectors would enhance the performance of both the single-photon and missing-momentum techniques.

On the contrary, no rate issues are expected in dump mode, as the beam and all secondary interaction are fully absorbed by the dump itself. In this configuration, luminosity can be increased by a factor of approximately 100 by both raising the number of incoming positrons and increasing the target thickness to tens of centimeters, taking advantage of interactions from secondary particles produced in electromagnetic showers.

In order to illustrate the potential of such an experimental configuration in searching for light dark sectors, we will show the projected reach for well-motivated new physics models, following the classification proposed in Ref.~\cite{Antel:2023hkf}. We will consider a dark photon (DP) that interacts with the electromagnetic current via a coupling suppressed by a small kinetic mixing parameter $\varepsilon$. 
As a second candidate, we will consider axion-like particles (ALP), namely pseudo-scalars interacting with the photons via a coupling $g_{a\gamma}$,  
and with electrons via a coupling  $g_{ae}$.

\subsection{Dark Photon production in positron-on-target collisions}
\label{Sec:DPproduction}

The  interaction of a DP  $A'_\mu$ 
with a fermion $\psi_f$ is described by the following 
Lagrangian:
\begin{equation}
    \mathcal{L}_{A'} \supset -i g_V \bar{\psi}_f \gamma^\mu \psi_f A'_\mu\,.
\label{eq:Aprime}
\end{equation}
This interaction holds also for DP  models where 
the DP    interacts with a SM fermion $\psi_f$ via 
a kinetic mixing parameter $\varepsilon  = g_V /e$.

\begin{figure}[t]
\begin{center}
\resizebox{1.\linewidth}{!}{%
   \includegraphics{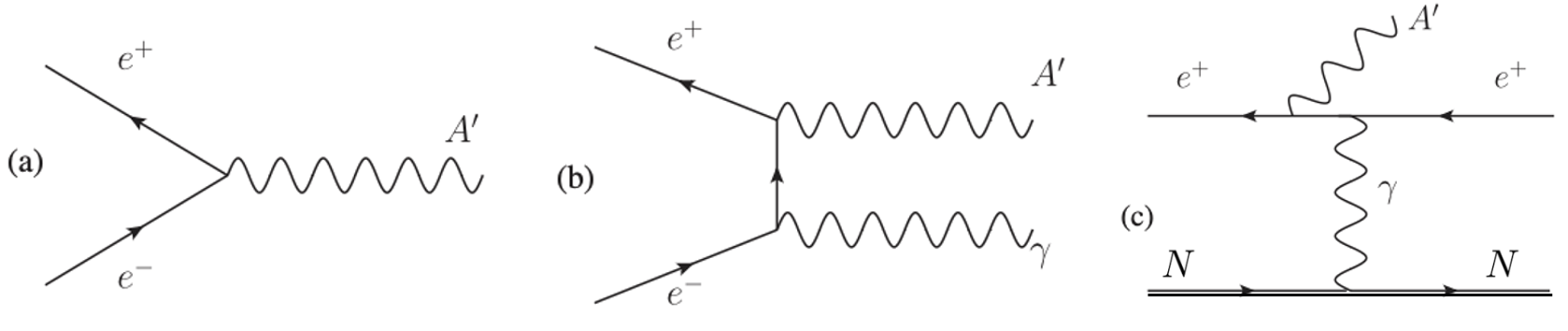}
 }
\caption{A$^{\prime}$ production mechanisms in electron-positron collisions~\cite{Nardi:2018cxi}}.
\label{fig:AProd}    
\end{center}
\end{figure}

In positron collisions on fixed targets, 
the dominant production modes for a DP  $A'$ are:
\begin{enumerate}
    \item[a)] Resonant annihilation:   $e^+e^- \to A'$
    \item[b)] Associated production:   $e^+e^- \to\gamma A'$
    \item[c)] Bremsstrahlung: $e^+N \to e^+ N A' $
\end{enumerate}
The corresponding Feynman diagrams are shown in Fig.~\ref{fig:AProd}.

The  first two production processes require positron beams,   
and offer an important advantage in terms of production rates compared to electrons beams, for DP  masses around the resonant value $m_{\textrm{res}} =\sqrt{2 m_e E_{Beam}} \sim 277 \, \textrm{MeV}$.
The resonant production mechanism has the highest cross-section due to a single electromagnetic vertex
and Breit–Wigner enhancement.
Unfortunately, this occurs only within a narrow mass range, very close to the $\sqrt{s}$ of the collision. At lowest order, this process yields a  cross section  that, away from the resonant c.m. energy, is proportional to the   $A'$ width $\Gamma_{A'}$,  which is strongly suppressed by the small coupling constant $\varepsilon$~\cite{Nardi:2018cxi}

\begin{equation}
\sigma_{res} = \sigma_{\rm peak}\frac{\Gamma^2_{A'}/4}{(\sqrt{s}- m_{A'})^2+\Gamma_{A'}^2/4}
\label{eqn:ResProd}
\end{equation}
where $s$ is the invariant mass squared, and $\sigma_{\rm peak}= 12 \pi/m_{A'}^2$~\cite{Nardi:2018cxi}. 
While the extremely narrow width of the DP  makes this distribution strongly peaked around the resonant mass $m_{\textrm{res}}$, two main effects  concur to widen significantly the mass range where this production mechanism dominates.
First, the RMS   beam energy spread  $ \sigma_\textrm{Beam}/E_{\textrm{Beam}} \simeq 1\% $ is modelled using a Gaussian distribution, then further convolved with the 
 differential track lengths $\frac{\partial T_{e}}{\partial E}$ representing the
 total length (in unit of radiation length) 
of the material  traversed by 
all $e^\pm$ of a given energy $E$  present in the shower. 
 We use analytical expressions derived in~\cite{Tsai:1966js}:
\begin{align}
\label{eq:tracklenghtTsai}
 \frac{\partial T_{e}}{\partial E} = \int_{0}^{t_{tar}} \! dt \ 
 I_{e} (t,E), \qquad
 \\
    I_{e} (t,E) = \displaystyle \frac{1}{E}\frac{\left[\ln(E_0/E))\right]^{4t/3-1}}{\Gamma \!\left(4t/3\right) }, 
\end{align}
where $t_{tar}$ is the target length in unit of radiation length and $E_0$ is the initial positron energy before entering the target. 

Second, the atomic electrons in the target are nor free nor at rest.
Their spatial localization implies  a certain momentum distribution,
which in turn can alter significantly  the CoM energy of the $e^+ e^-$ system.
We model this effect following Ref.~\cite{Arias-Aragon:2024qji}. 

Finally, in the associated production case, the DP  is produced alongside a Standard Model  (SM) photon. Due to the  initial-state radiation (ISR) photon, which modifies the collision’s c.m. energy, the mass range for which associate production is effective is broader.
The accompanying photon can be used to tag the invisible decay of $A'$ and enables the measurement of its mass as the missing mass, calculated as 
$M^2_{A'} = (p^\mu_{e+} - p^\mu_{e-} - p^\mu_{\gamma})^2$.
 
The total cross section as a function of the $A'$ mass $m_{A'}$ and of the coupling $\varepsilon$ reads~\cite{Marsicano:2018krp} 
\begin{equation}
\begin{aligned}
\sigma_{\mathrm{ass}}(m_{A'},\varepsilon)
&= \frac{8\pi\alpha^2 \varepsilon^2}{s}
\Bigg[
\left(\frac{s-m_{A'}^2}{2s}
+ \frac{m_{A'}^2}{s-m_{A'}^2}
\right)
\\
&\quad\times
\log\!\frac{s}{m_e^2}
- \frac{s-m_{A'}^2}{2s}\Bigg]\,.
\end{aligned}
\label{eq:AssProd}
\end{equation}

where the c.m. energy of the collision is 
$\sqrt{s}$. 
The cross-section significantly increases when the mass of the particle is close to $s$, approaching the resonant value in the limit of vanishing photon energy. In practice, we merge both processes by keeping only the dominant one close to the resonant energy.
The acceptance for this production process is high in NA62 due the low photon emission angle caused by the high boost factor. 

Finally, the dark bremsstrahlung process is similar to the SM bremsstrahlung one. In the so-called Weizsacker-William approximation~\cite{Tsai:1986tx,Bjorken:2009mm}, we can write the integrated cross-section as
\begin{align}
\label{eq:DPbrem}
\sigma \approx \frac{4}{3} \frac{\alpha_{\rm em}^3\varepsilon^2 \mathcal{F} \beta_{V}}{m_{V}^2} \, \log\left(\frac{1}{(1-x)_c }\right)\,, 
\end{align}
where $ (1-x)_c = \max\bigg( \frac{m_e^2}{m_{V}^2},\, \frac{m_{V}^2}{E_0^2}\bigg) $ and $\frac{\alpha_{\rm em} \mathcal{F}}{\pi}$ describes the effective photon flux (see, e.g.~\cite{Celentano:2020vtu}). In practice, we obtain 
this cross-section numerically 
by means of the \texttt{MadGraph5\_aMC}$@$\texttt{NLO}  simulation framework~\cite{Alwall:2014hca}, following the implementation of~\cite{Celentano:2020vtu},  using an effective $N N \gamma$ interaction between the nuclei and the photon with form factor $G_2$. The form factor for a nucleus $(Z,A)$ is a function of the exchanged photon virtuality $t$  and $G_2(t) = G_2^{el} + G_2^{in} $ is defined by:
\begin{align}\nonumber
	G_2^{el}&=\left(\frac{a^2 t}{1 + a^2 t}\right)^2 \left(\frac{1}{1 +
	t/d}\right)^2 Z^2\,, \\
	G_2^{in}&=\left(\frac{{a^\prime}^2 t}{1+{a^\prime}^2 t}\right)^2 \left(\frac{1 +
		\frac{t}{4m_p^2}(\mu_p^2 - 1)}{\left(1 +
		\frac{t}{0.71\,\mathrm{GeV}^2}\right)^4}\right) Z \, ,
\end{align}
with $\mu_p = 2.79$,  $m_p = 0.938$ GeV the proton mass, $  a \equiv 111 \frac{1}{m_e Z^{1/3}} ,\ \  a' \equiv 773 \frac{1}{m_e Z^{2/3}} $ and $ d~=~0.164 ~\textrm{ GeV}^2 A^{-2/3} $.

\subsection{Axion Like Particles production in positron-on-target collisions}

A thorough study of production of Axion-Like Particles (ALPs) in positron-on-target collisions can be found  in~\cite{Darme:2020sjf}.
In general, ALPs can interact both with photons and electrons via the couplings $g_{a\gamma}$ and $g_{ae}$ respectively, 
as described by the following Lagrangian:
\begin{equation}
    \mathcal{L}_a\supset \frac{g_{a\gamma}}{4}aF^{\mu\nu}\tilde{F}_{\mu\nu}+\frac{g_{ae}}{2}\partial_\mu a \, \bar{e}\gamma^\mu\gamma_5e,
\end{equation}
where $\tilde{F}_{\mu\nu}=\frac{1}{2}\varepsilon_{\mu\nu\rho\sigma}F^{\rho\sigma}$ with $\varepsilon_{1230}=+1$ is the photon dual field strength tensor.
Depending on the assumption on the dominant coupling of the ALPs to the SM, two different phenomenology of production and decay are obtained:
\begin{enumerate}
\item ALPs with dominant electron coupling ($g_{ae}\neq 0$ and $g_{a\gamma}\approx 0$)
\item ALPs with dominant photon coupling    ($g_{a\gamma} \neq 0$ and $ g_{ae} \approx 0$)
\end{enumerate}

\subsubsection{ALPs with dominant electron coupling}
The main production mechanisms for ALPs with dominant coupling to electrons $g_{ae}\neq 0$ and $g_{a\gamma}\approx 0$ are identical to the DP  one shown in Fig.~\ref{fig:AProd}:

\begin{enumerate}
    \item[a)] Resonant annihilation:   $e^+e^- \to a$,
    \item[b)] Associated production:   $e^+e^- \to a \gamma$,
    \item[c)] ALP bremsstrahlung : $e^+N \to e^+ N a $.
\end{enumerate}
For this reason the detection techniques and the resulting limits can largely be obtained with the same procedure just by rescaling for the different production and decay rates. The cross section for resonant annihilation can be written straightforwardly after accounting for the beam energy spread as:

\begin{equation}
\begin{aligned}
\sigma_{\mathrm{res}}(m_a,g_{a\gamma})
&= \frac{\sqrt{\pi}\,|g_{a\gamma}|^2 m_e}{4\sqrt{2}\,\sigma_E}
   e^{-\frac{\left(E_{\mathrm{res}}(m_a)-E_B\right)^2}{2\sigma_E^2}},
\\
&\quad
E_{\mathrm{res}} = \frac{m_a^2}{2m_e} - m_e .
\end{aligned}
\label{eq:res}
\end{equation}

The ALPs associated production cross section in the hypothesis of dominant electron coupling reads~\cite{Darme:2020sjf}
\begin{equation}
\begin{aligned}
\sigma_{\mathrm{ass}}(m_a,g_{ae})
&= \alpha_{\mathrm{em}}\, g_{ae}^2 m_e^2
\\
&\quad\times
\frac{
-2 m_a^2 \beta s
+ \left(s^2 + m_a^4 - 4 m_a^2 m_e^2\right)
\log\!\frac{1+\beta}{1-\beta}
}{
2 (s - m_a^2) s^2 \beta^2
}\,.
\end{aligned}
\label{eqn:ALPsProd}
\end{equation}

where $\beta=\sqrt{1-\frac{4m_e^2}{s}}$.

The cross section for ALP bremsstrahlung is obtained numerically using the same framework as for the DP  case~\cite{Darme:2020sjf}. A simple order-of-magnitude estimate; accurate at around $10\%$ can be straightforwardly obtained by dividing by two the DP  result in Eq.~\eqref{eq:DPbrem}, along with the replacement $\varepsilon \to g_{ae} m_e /e$.

\subsubsection*{ALPs with dominant photon coupling}

The scenario is markedly different for models where ALPs exhibit dominant photon couplings, with $g_{a\gamma} \neq 0$ and $ g_{ae} \approx 0$. In this case the following production mechanisms are available :

\begin{enumerate}
    \item[a)] Resonant annihilation:      $e^+e^- \to a$
    \item[b)] ALP-strahlung:   $e^+e^- \to\gamma a$
    \item[c)] Photon fusion: $e^+e^- \to e^+ e^- a$ or $e^+N \to e^+ N a$
    \item[d)] Primakoff: $\gamma N \to a N$
\end{enumerate}

\begin{figure}[t]
\begin{center}
\resizebox{0.9\linewidth}{!}{%
\includegraphics{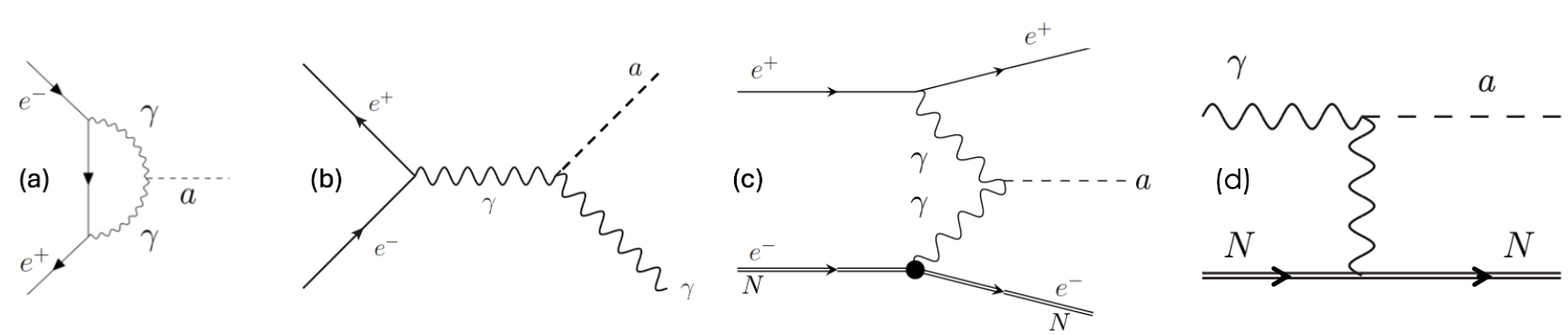}
}
\caption{ALPs with dominant coupling to photons production mechanisms in positron-on-target collisions.~\cite{Dolan:2017osp}}
\label{fig:ALPsPhoton}    
\end{center}
\end{figure}

Among the various production mechanisms, ALP-strahlung, photon fusion, and Primakoff (see diagrams (b), (c), and (d) in Fig.~\ref{fig:ALPsPhoton}) are the most significant in positron-on-target collisions.\footnote{When the tree-level electron coupling vanishes, resonant production  $e^+e^-\to a$  can still proceed via the loop diagram in Fig.~\ref{fig:ALPsPhoton}\;(a), and can be particularly relevant when the beam energy 
is tuned to the resonant value $E_{res}$ and the beam energy spread $\sigma_E$ is small, see Eq.~(\ref{eq:res}). However,  the effect of atomic electron momentum distribution 
that produces a smearing of the actual  
c.m. energy in the collision  
renders ineffective reducing the beam energy spread 
below $\sim 1\%$~\cite{Nardi:2018cxi}. This effect is especially relevant in high $Z$ materials as inner shell electrons can reach relativistic 
velocities~\cite{Arias-Aragon:2024qji,Arias-Aragon:2024gpm}.
} These processes lead to distinct final states: ALP-strahlung produces a single photon and an ALP, while photon fusion results in an $e^+e^-$ pair along with an ALP. 
Primakoff production (d) relies on secondary photons generated in the electromagnetic shower and is particularly relevant in the thick-target scenario. We estimated both on-shell (Primakoff) and off-shell (photon-fusion) processes with \texttt{MadGraph5\_aMC}$@$\texttt{NLO}, using the same nucleus form factor as for DP  bremsstrhalung. In both cases, if the momentum transferred to the nucleus by the virtual photon is smaller than the inverse of the nuclear size, the incoming positron interacts coherently with the whole nucleus, resulting in a cross-section enhancement by a factor of $Z^2$, where $Z$ represents the atomic number of the nucleus~\cite{Gao:2024rgl}. Primakoff production strongly dominates due to the presence of multiple photons in the electromagnetic shower for thick targets. We additionally found an excellent agreement with existing analytical expressions~\cite{Dobrich:2015jyk} for the $\gamma N \to a N$ process cross-section.

Additionally, even in the case of a vanishing axion-electron coupling at tree level, direct annihilation is possible via a loop-induced coupling (see diagram $(a)$ in Fig.~\ref{fig:ALPsPhoton}). In this case, we  
consider the following effective ALP-electron coupling 
for ALP production~\cite{Chala:2020wvs,Arias-Aragon:2022iwl}

\begin{equation}
\begin{aligned}
g_{ae}^{\mathrm{eff}}
&= \alpha g_{a\gamma} \frac{i}{4\pi m_e^2}
\Bigl(3A_0\!\left(m_e^2\right)
\\
&\quad
+ m_e^2\!\left(2 - m_a^2 C_0\!\left(m_a^2, m_e^2, m_e^2, 0, 0, m_e^2
\right)
\right)
\Bigr)
\,,
\end{aligned}
\end{equation}

where $A_0$ and $C_0$ are the usual Passarino-Veltman functions that can be implemented using numerical packages like \texttt{Looptools}~\cite{vanOldenborgh:1989wn,Hahn:1998yk}. 

Accounting for the beam spread, the cross section for resonant $e^+e^-\rightarrow a$ production is written as before, simply accounting for the effective coupling
\begin{equation}
    \sigma_{res}(m_a,g_{a\gamma})=\frac{\sqrt{\pi}\left|g_{ae}^{eff}\left(m_a,g_{a\gamma}\right)\right|^2m_e}{4\sqrt{2}\sigma_E}e^{-\frac{\left(E_{res}\left(m_a\right)-E_{B}\right)^2}{2\sigma_E^2}},
\end{equation}
while the associate production this time takes the following form~\cite{Darme:2020sjf}:
\begin{equation}
    \sigma_{ass}\left(m_a,g_{a\gamma}\right)=\alpha g_{a\gamma}^2\frac{\left(s+2m_e^2\right)\left(s-m_a^2\right)^3}{24\beta s^4}\,.
\end{equation}

Notably, when the tree-level electron coupling vanishes, the loop-induced coupling causes resonant production to dominate over ALP-strahlung within a narrow mass range, determined by the beam spread \(\sigma_{\text{Beam}}/E_{\text{Beam}} \approx 1\%\). This range would expand further if electron motion were considered, as relativistic electrons in inner shells increase the available c.m. energy.

\subsection{Dark sector particles decay modes}

If the dark sector is composed of more than one new particle, its internal structure will play an important role in fixing the relevant experimental signatures. The lightest dark sector particle, if unstable, will for instance typically decays back into SM particles, while  heavier particles might experience semi-visible or fully-invisible decays patterns. Given that the dominant production modes for dark sector in NA62e+ are via lepton- or photon-driven processes, we will focus in this work on visible decays of the new particle $X$ into lepton pairs for vector particles and photon pairs for ALPs. 

Conversely, if additional light stable or very long-lived dark sector particles $\chi$ exist with masses below half the $X$ mass, the dark sector particle will predominantly decay invisibly into $\chi\bar{\chi}$, the so-called invisible decay scenario.
Different combinations of production and decay modes lead to distinct experimental signatures. In the next section, we will show how NA62, with its versatile detector, can probe a large range of possibilities.

\section{Invisible decay search techniques at NA62e+}

\begin{figure}[ht]
\begin{center}
\resizebox{0.85\linewidth}{!}{%
   \includegraphics{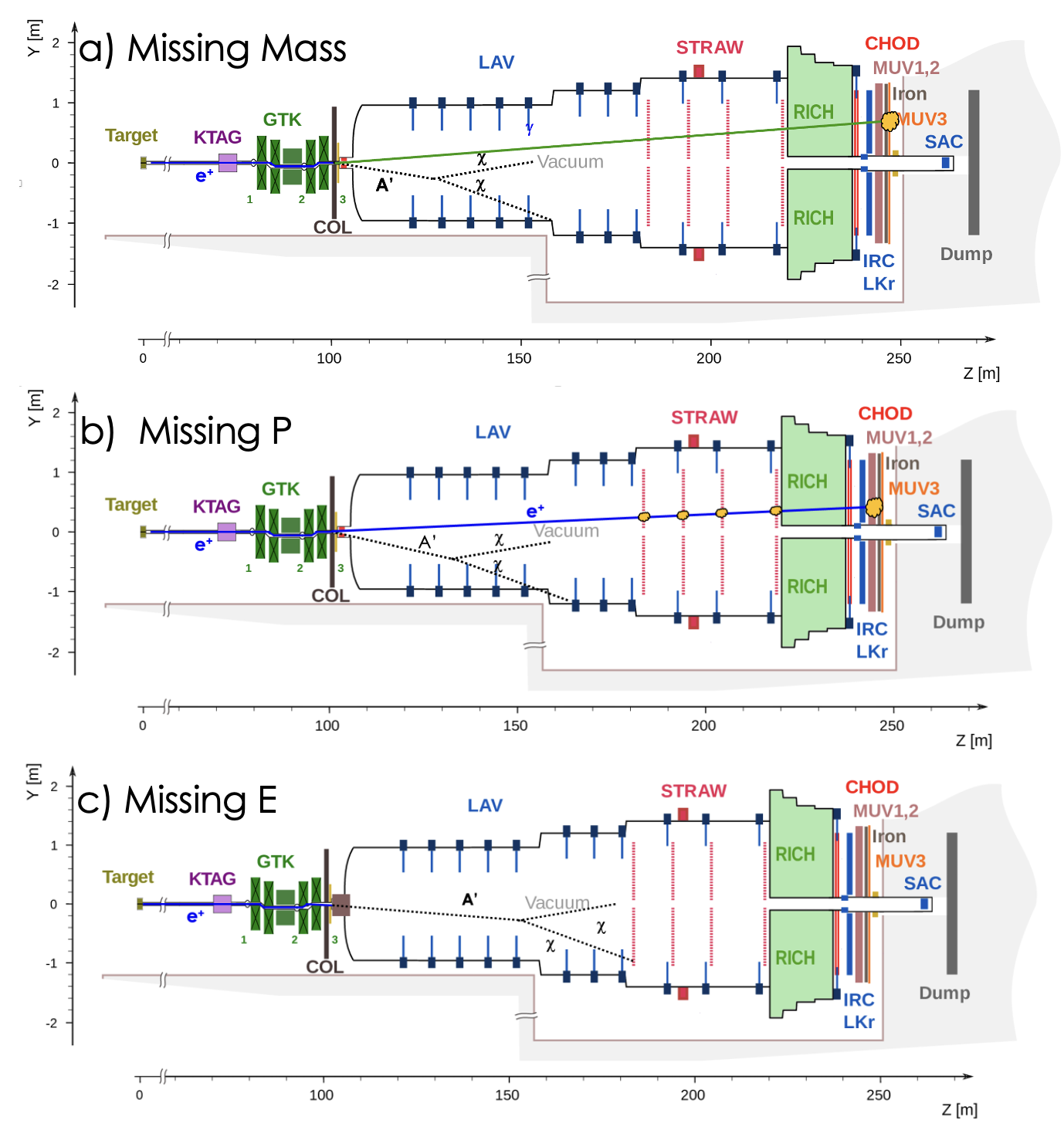}
}
\caption{NA62e$^+$'s invisible decay search techniques illustrated in the case of a DP  $A'$. a) Missing mass b) Missing momentum c) Missing energy.
} 
\label{fig:InvTech}    
\end{center}
\end{figure}

In this section, we assume that any dark sector particle $X$ produced in positron-on-target collisions is either stable or primarily decays into lighter dark sector states $\chi\bar{\chi}$, which escape detection. In both cases, the final states are characterized by significant missing energy.
This missing energy can be identified using different techniques, depending on the production mechanism and target thickness. Figure~\ref{fig:InvTech} illustrates how the NA62 detector could be used to implement these techniques.

\subsection{Missing mass: $e^+e^- \to \gamma X $ }

Mono-photon searches exploit associated production diagram b) in Fig. \ref{fig:AProd}, measuring the SM photon's four-momentum with the NA62 LKr calorimeter to infer the presence of an invisible particle. The NA62 Giga Tracker spectrometer precisely measures the incoming positron's energy and direction, allowing for a precise determination of the missing particle's invariant mass, 
$M^2_{X}=(p^\mu_{e+}-p^\mu_{e-}-p^\mu_{\gamma})^2$. 
A schematic view of the technique is given in Fig.~\ref{fig:InvTech} a).
NA62's excellent photon veto system enables nearly background-free searches, making it ideal for detecting DPs, ALPs, or very long-lived particles that decay invisibly or escape detection.

The search technique relies on identifying “bumps” in the reconstructed missing-mass spectrum $M_X^2$. Owing to the excellent $M_X^2$ resolution of $\sim$ 1 MeV achieved by NA62~\cite{NA62:2019meo}, approximately hundred different mass hypotheses can be tested.

The main backgrounds for the single photon searches are $e^+e^- \to \gamma \gamma$ with a missing photon, and the positron bremsstrahlung with a missing positron.

In the bremsstrahlung case, the high Lorentz boost factor $\gamma$ causes almost all recoil photons to be emitted collinearly with the incoming positron, with a typical angle of $1/\gamma \sim 6~\mu\text{rad}$. As a result, the recoil photon will very rarely enter the NA62e+ calorimeter acceptance and therefore cannot mimic a signal-like final state.  Even in the cases in which the recoil photon is detected in the calorimeter, the outgoing positron will be captured by the NA62e+ charged veto system with a rejection capability of the order of $1\times 10^{-5}$.

According to our CalcHEP simulation, the cross section $\sigma(e^+e^- \to \gamma\gamma)$ is $\sim 2\times10^7$ pb. Moreover, there is no kinematical configuration in which one photon enters the LKr calorimeter acceptance while the other escapes the 50 mrad coverage of the NA62e+ photon veto detectors.
Using the luminosity calculated in Eq.~\ref{eqn:NA62Lumi}, the expected background can be estimated as follows:

\begin{equation}
N^{2\gamma}_{BG}= \sigma(e^+e^- \to \gamma\gamma) \times \text{NA62}_{\text{lumi}}^{Si,W}\times \varepsilon_{\text{veto}}\times Acc_{1\gamma}/N_{\text{Mass}}
\end{equation}
where $\varepsilon_{\text{veto}}$ represents the NA62 photon veto inefficiency, and $N_{Mass}$ is the number of mass hypothesis tested. Using reasonable parameter values, namely $\varepsilon_{\text{veto}} = 2\times 10^{-5}$, a one-photon acceptance $Acc_{1\gamma} = 0.1$, and $N_{\text{Mass}} = 150$, we expect approximately one background event per $2\times 10^{14}$ $e^+$OT.

To compute the sensitivity of NA62e+ to the $A'$ and ALPs invisible decays we used Eq.~\ref{eq:AssProd} and Eq.~\ref{eqn:ALPsProd} for the cross section ($CS_{A',a}$) $e^+e^-\to \gamma (A',a)$ for each mass values from 0 up to the kinematic limit $M_{A',a}^{Max}=\sqrt{2\times E_{Beam} \times m_e}\sim$ 277 MeV.
From the obtained CS values we estimated the 90\% confidence level upper limit on the coupling $\varepsilon$, $g_{ae}$ as the Single Event Sensitivity (SES) for $2\times10^{14}$ e$^+$OT using the following formula:
\begin{equation}
UL(\varepsilon, g_{ae})=\sqrt{\frac{2.3}{Acc_{A',a}\times CS_{A',a}\times \text{NA62}_\text{lumi}^{Si,W}}}\,.
\label{eqn:ULFormula}
\end{equation}
The signal acceptance $Acc_{A',a}$ was determined using the \texttt{CalcHEP} Monte Carlo generator~\cite{Belyaev:2012qa} for different mass values. 
The $A'$ model was implemented into \texttt{CalcHEP} and we applied an angular selection cut of 1 mrad $< \theta_{\gamma} <$ 8.5 mrad to the emitted recoil photon to ensure it would be detected by the NA62 LKr calorimeter. As a result, we found that $\sim$15\% of the events are accepted by the calorimeter with a slightly increasing trend for higher masses.
Additional acceptance gains can be explored using small-angle calorimeters to detect the recoil photon, provided the trigger rates and detector performance are adequate. 

\begin{figure}[h]
\begin{center}
  \centering
  \resizebox{\linewidth}{!}{%
    \includegraphics{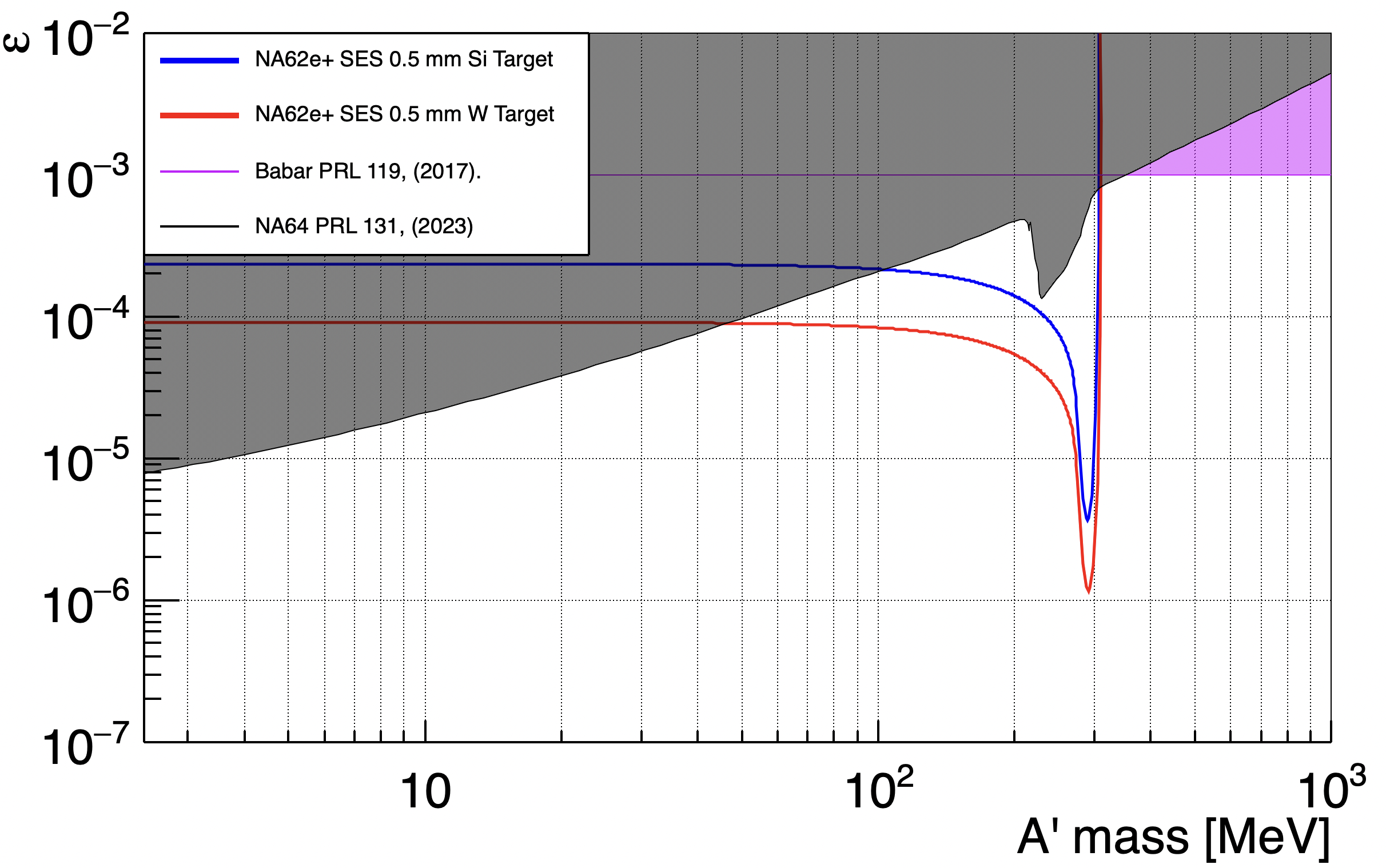}
  }
  \caption{Expected NA62e+ sensitivity to $A'$ invisible decays in different scenarios. Blue line Silicon and red line tungsten target in single photon mode. 
  Filled grey area represents NA64~\cite{NA64:2023wbi} and filled\ green one the Babar~\cite{BaBar:2017tiz} limits. 
  }
  \label{fig:AExclInv}
\end{center}
\end{figure}

The single event sensitivity to $A'$ invisible decays expected in NA62e+ are shown in Fig.~\ref{fig:AExclInv}.
The filled areas represent existing limits from NA64~\cite{NA64:2023wbi} (grey) and Babar~\cite{BaBar:2017tiz} (green).  
The solid curves represent the expected Single Event Sensitivity (SES) for the mono photon analysis at NA62e+ with $2 \times 10^{14}$ e$^+$OT, using 500 $\mu$m thick tungsten (red) and silicon (blue) targets, respectively. 

\begin{figure}[h]
  \begin{center}
  \resizebox{\linewidth}{!}{%
    \includegraphics{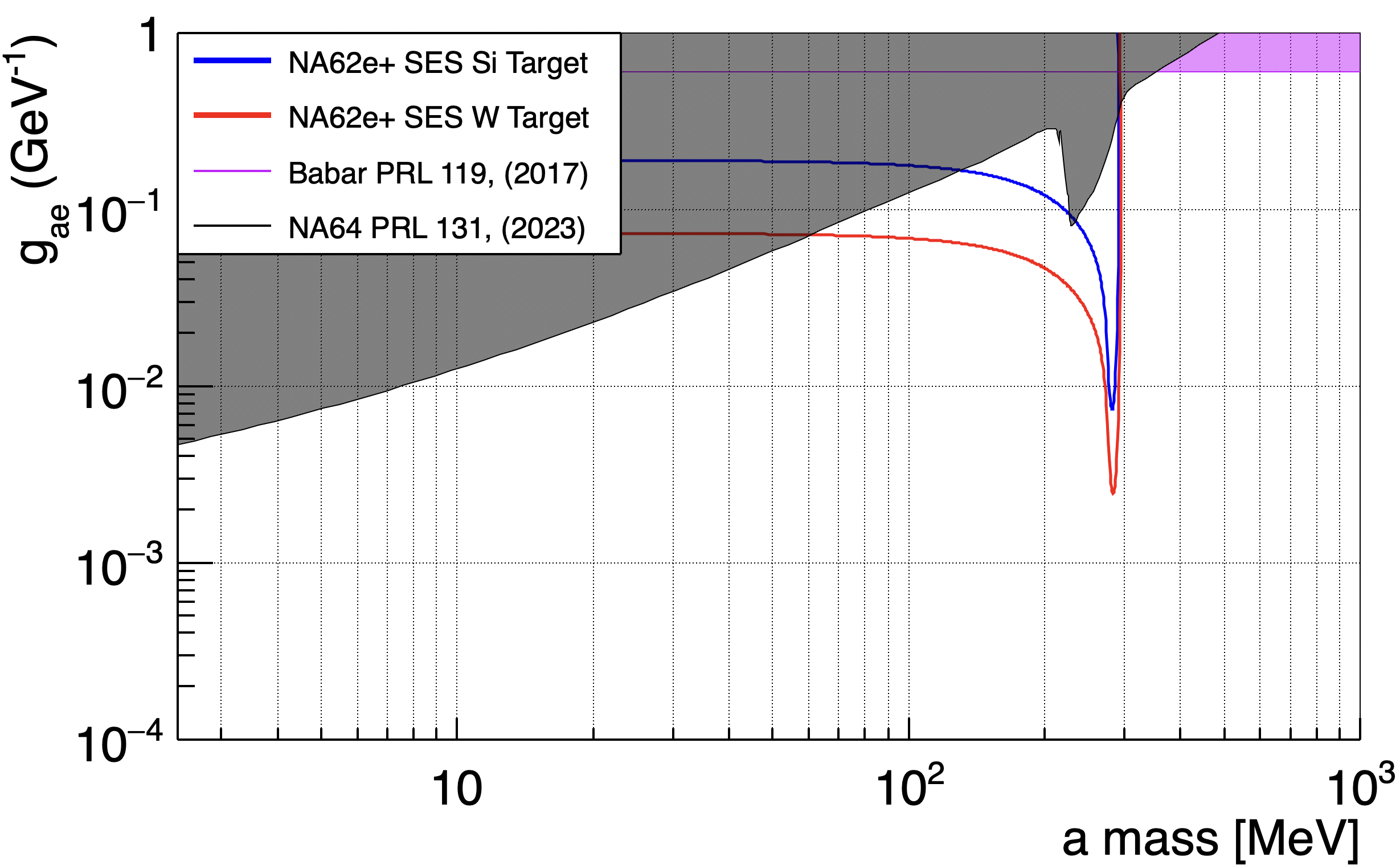}
  }
  \caption{Expected NA62e+ sensitivity to ALPs ($g_{ae}$) invisible decays in different scenarios. The blue (red) line represents a silicon (tungsten) target in single photon mode. 
  The shaded gray area shows the NA64 limits~\cite{NA64:2023wbi}, and the green one the BaBar limits~\cite{BaBar:2017tiz}.} 
  \label{fig:ALPExclInv}
\end{center}
\end{figure}

Similar limits are obtained for invisible decays of ALPs with dominant coupling to electrons shown in Fig.~\ref{fig:ALPExclInv}. We used as in the DP  case the zero background approximation and a 15\% acceptance, estimated using \texttt{CalcHEP}~\cite{Belyaev:2012qa}. 

Profiting by the higher production rates provided by the associated production at high mass compared to $A'$ bremsstrahlung, NA62e+ can improve up to 2 orders of magnitude the present limits by NA64 in the mass region 150--277 MeV for both $A'$ and ALPs invisible decays.
Moreover, several different regions of parameters space can be probed by simply changing the positron beam momentum. Pushing the beam energy to 150 GeV, as in the NA64 case, the kinematical limit for the single photon analysis will reach the 400 MeV region. The lower positron flux at higher energy can be compensated by increasing the target thickness. 

Finally, the motion of atomic electrons, particularly in the tungsten target case, can significantly enhance the $A'$ production rates to masses beyond the kinematic limit of the electron-at-rest approximation~\cite{Arias-Aragon:2024qji}.
However, its effect on the missing mass resolution has not yet been calculated, and we therefore do not include it in the projected reach shown in Figs.~\ref{fig:AExclInv} and~\ref{fig:ALPExclInv}. We nonetheless note that  it is expected to be negligible in Si compared to W target.

\subsection{Missing momentum: $e^+N \to e^+ N X$}

The missing momentum experiment exploits the bremsstrahlung $X$ production mechanism. In this process, the dark sector particle is emitted from a primary positron after interacting with a target nucleus. The $X$ subsequently decays into invisible particles, leading to a significant reduction in the final-state positron's momentum compared to its initial value.
The measurement of the $e^+$ momentum downstream of the target reveals a significant difference from the original beam momentum. A schematic view of this technique is shown in Fig.~\ref{fig:InvTech} b).  
This technique, proposed for the first time for the LDMX experiment~\cite{Izaguirre:2014bca}, is considered the one with lowest background and has not been used until now. 

To conduct such an experiment, both an upstream tracker before the target and a recoil tracker after the target are essential.
The NA62 experiment is designed as a missing momentum experiment. Notably, the decay  $K^+ \to \pi^+ \nu \bar{\nu}$ involves a missing momentum final state, closely resembling the process  $e^+ \to e^+ A^{\prime} \to e^+ \chi \bar{\chi}$. Compared to LDMX, NA62 offers the capability to reconstruct the invariant missing mass and features a very precise particle identification (PID) system, which enhances background rejection, particularly against hadronic final states. The high energy of the incoming positrons assure the complete angular coverage of photon veto detectors for SM backgrounds. 
The capability of obtaining very high rejection factor  
on extra particles has been demonstrated in the recent measurement of the $BR(K^+ \to \pi^+ \nu \bar{\nu})$~\cite{PnnRun1Paper}.

In the NA62 experiment, the GTK system serves as the upstream tracker, providing a 0.2\% measurement of the incoming positron's momentum. 
The momentum of the positron after the interaction with the target is measured by the NA62 magnetic spectrometer with a relative resolution parameterized as~\cite{NA62:2017rwk}:

$$\sigma(p)/p = 0.48\% \oplus 0.009\%\times p$$ with $p$ in GeV/c.

\begin{figure}[h]
    \centering
    \includegraphics[width=0.9\linewidth]{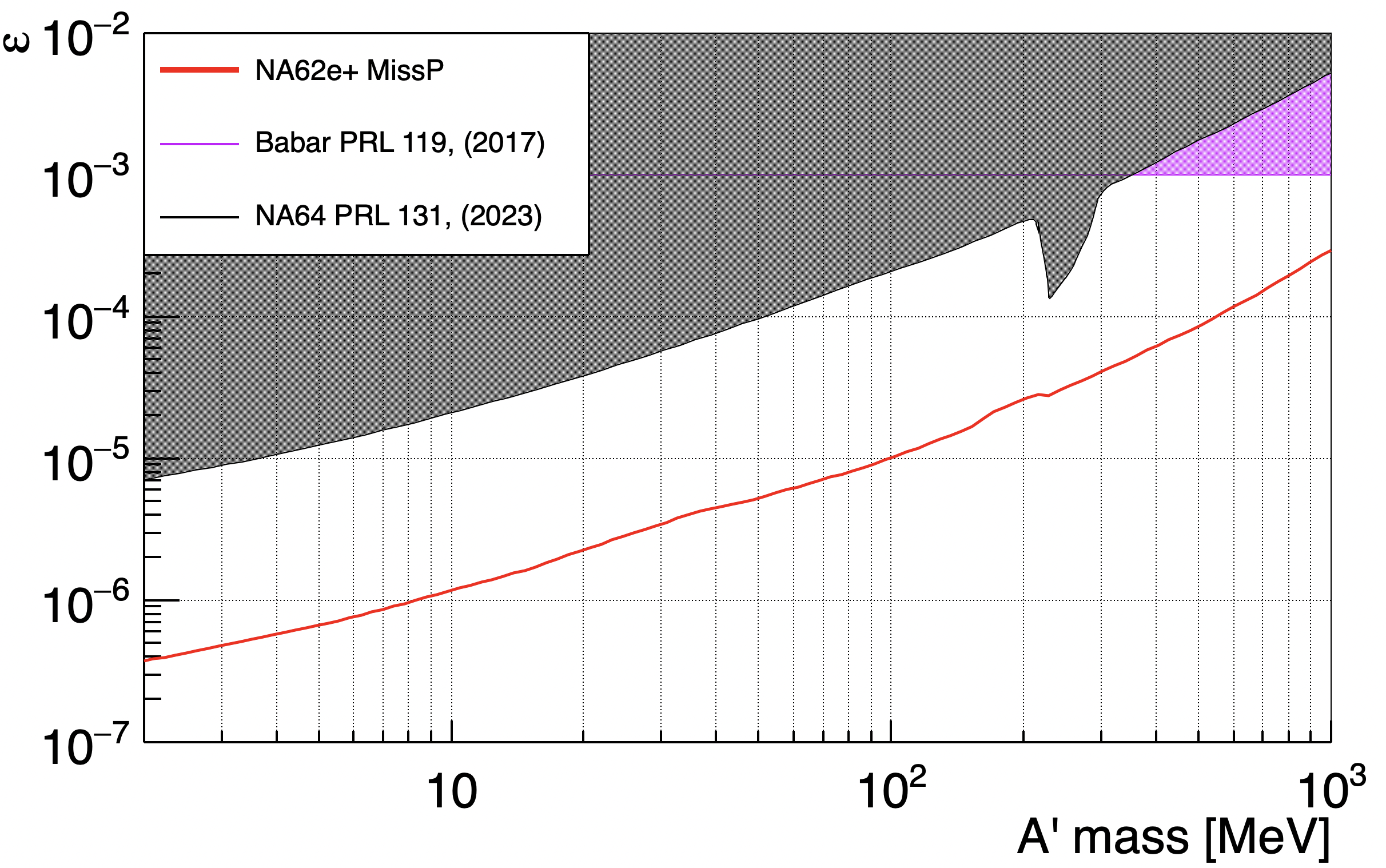}
    \caption{NA62e+ missing momentum sensitivity to invisible $A'$ decays (red) compared to existing limits.}
    \label{fig:InvMissP}
\end{figure}

The NA62e+ sensitivity shown in Fig.\ref{fig:InvMissP} has been obtained assuming $0$ background and applying a conservative missing momentum cut of 50\%.
 
Detailed background studies have been conducted for the LDMX experiment at SLAC \cite{CraigGroup:2025rwc}, demonstrating that no irreducible BG sources exist up to $\sim 10^{16} e^+$OT for a missing momentum of 50\%. 
Therefore, we believe it is safe to assume that the same BG-free hypothesis holds for $\sim 10^{14} e^+$OT. In addition the tenfold higher beam energy of the NA62e+ experiment provides a clear advantage in both veto efficiency and calorimeter resolution.

The limit was calculated using Eq.~\ref{eqn:ULFormula} in which $Acc$ and CS have been obtained using \texttt{MadGraph5\_aMC}$@$\texttt{NLO}~\cite{Alwall:2014hca} as described in Sec.~\ref{Sec:DPproduction}. 
The acceptance was calculated ensuring that the recoils positron lies into the spectrometer acceptance (1 mrad$<\theta<$8.5 mrad). A very similar acceptance to the case of mono-photon $\sim$15\% has been obtained. 

\subsection{Missing energy: $e^+Z \to e^+ \cancel{E} $}
\label{par:MissE}

In the missing energy technique an active thick target absorbs and measures the incoming positron. If the electromagnetic shower a dark sector particle $X$ is produced, the measured energy differs significantly from the incoming positron momentum.  
A missing energy search can potentially be performed at NA62e+ installing an active target downstream the GTK3 station, shown in brown in Fig.~\ref{fig:InvTech} c), and using the rest of the NA62 detector as a global veto system.

\begin{figure}[h]
  \centering
  \resizebox{\linewidth}{!}{%
    \includegraphics{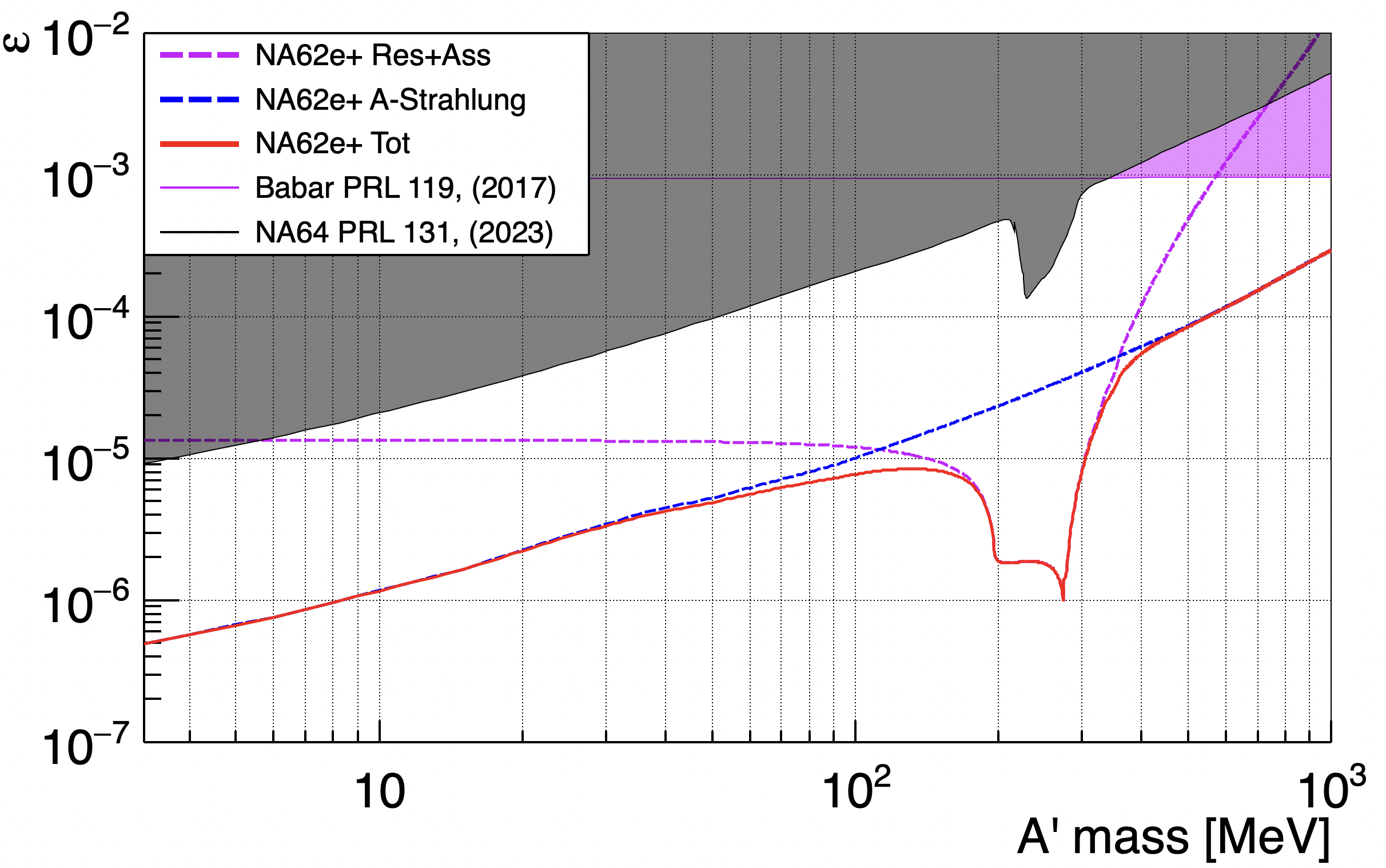}
  }
  \caption{NA62e+ missing energy sensitivity to invisible $A'$ decays (red).}
  \label{fig:InvMissE}
\end{figure}

The NA64 experiment has extensively demonstrated the effectiveness of the missing energy technique using an electron beam on target. From 2016 to 2022, NA64 accumulated $9.37 \times 10^{11}$ 100 GeV electrons on target, achieving the most stringent limits on invisible DP  decays~\cite{NA64:2023wbi}. A similar search using positrons, as proposed in the POKER project, has also been undertaken by the NA64 collaboration. However, to date, only a limited amount of data, $1 \times 10^{10}$ positrons on target at 100 GeV, has been collected in this mode~\cite{NA64:2023ehh}.
In positron dump mode, NA62e+ has the potential to accumulate more than 1000 times the number of positron collisions compared to NA64\cite{NA64:2023ehh}, leveraging the high proton flux enabled by the North Area BDF upgrades. 
With minor interventions, a 50 cm long active target system, similar to POKER calorimeter~\cite{Antonov:2024glv}, can be installed in the NA62 experiment, downstream the GTK region, enabling missing energy measurements. This configuration will allow profiting from all the three production mechanisms provided by positron beams leading to the maximum sensitivity to invisible decays. The achievable exclusion limits with $2\times 10^{14}$ e$^+$OT, assuming no background, are shown in Fig.~\ref{fig:InvMissE} for $A^\prime$ searches.
This assumption is supported by the absence of background events in the NA64 experiment up to $1\times 10^{12}$ $e^-$OT~\cite{NA64:2023wbi}, as well as by more recent studies performed for the LDMX experiment~\cite{LDMX:2025ixw}. In addition, the use of an homogeneous calorimeter can provide even better BG rejection compared to the sampling calorimeters used by NA64 and LDMX collaborations.

\begin{figure}[h]
  \centering
  \resizebox{\linewidth}{!}{%
    \includegraphics{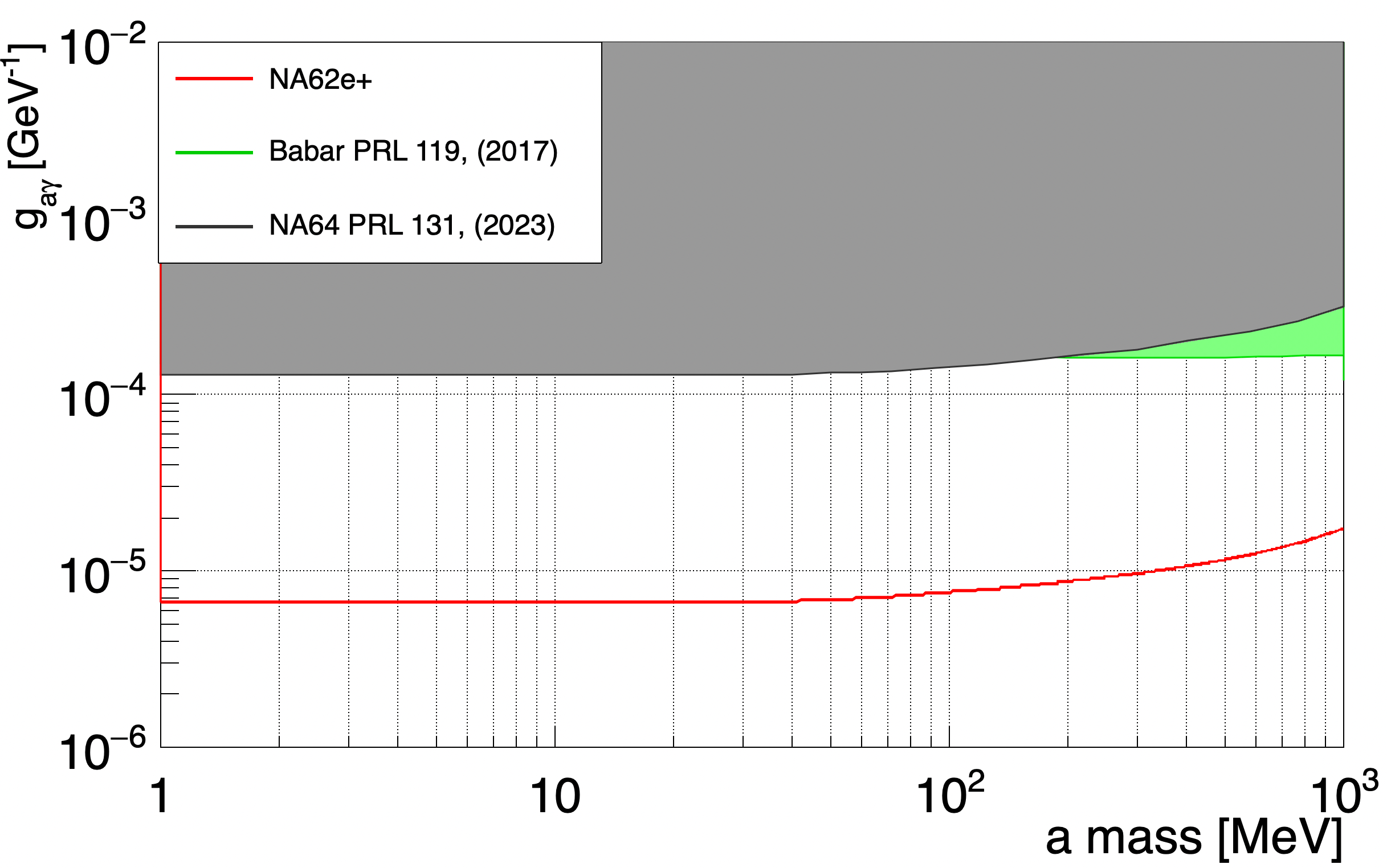}
  }
  \caption{NA62e+ missing energy sensitivity to invisible $a$ decays (red).}
  \label{fig:InvAlps_gg}
\end{figure}

The dashed blue line represents the contribution from 
$A'$-bremsstrahlung,
which is dominant at low masses. The production rate for this contribution has been obtained using \texttt{MadGraph5\_aMC}$@$\texttt{NLO} ~\cite{Alwall:2014hca}, accounting for the interaction in the whole target thickness and including a 50\% missing energy.
The violet dashed line shows the contribution of resonant and associated production, including interaction of the primary positrons and the contribution of secondary positrons generated in the electromagnetic shower obtained following the prescription in~\cite{Marsicano:2018glj}.
In this configuration, the limits from NA62e+ can surpass existing limits by more than one order of magnitude in a mass range from 1 MeV to 1 GeV, and by more than two orders of magnitude in the resonance-dominated region [200 MeV - 300 MeV].
According to~\cite{Schuster:2021mlr} additional sensitivity will be provided by invisible decays of photo-produced light mesons $\rho$, $\omega$, $\phi$. Mesons decay will improve the reach of
NA62e+ experiment for $m_{A'}>$ 500 MeV region in which $A'$-strahlung is weak, due to the $1/m_{A'}^{2}$ suppression of the cross section.

In the case of ALPs with photon-dominated coupling ($g_{a\gamma}$), the sensitivity was computed by considering ALP-strahlung and Primakoff production (diagrams (b) and (d) in Fig.~\ref{fig:ALPsPhoton}) while also accounting for a 50\% missing energy probability. The results, presented in Fig.~\ref{fig:InvAlps_gg}, show that sensitivity is primarily driven by Primakoff production across the entire mass range. The NA62e+ sensitivity (red) is compared to the exclusion limits from~\cite{Darme:2020sjf}, with the NA64 result (grey) rescaled to reflect the additional statistics collected by NA64~\cite{Andreev:2023xmj} relative to~\cite{Darme:2020sjf}. Notably, the NA62e+ sensitivity significantly surpasses that of NA64 and BaBar up to a few GeV in mass.
\section{Visible decay searches}

The visible decay model assumes that the produced dark sector particle $X$ decays mostly into SM particles. We further assume a $100\%$ branching ratios into SM leptons (or photon pairs for ALPs in the mass range we are interested in). 
Due to its small coupling to the SM leptons, the $X$ particle will exhibit a relatively long-lifetime that can result in decay vertices significantly displaced with respect to the production target thanks to the $\approx$1000 boost factor provided in NA62e+ by the 75 GeV beam energy. 

The NA62 detector is equipped with a spectrometer to measure the momenta and directions of charged particles and a LKr calorimeter to determine the energies of photons, enabling the full reconstruction of visible decays of dark sector particles. Additionally, a decay vertex position resolution $\approx$ $0.5$ m is achievable, allowing NA62 to distinguish long-lived particle decays from SM background.
\begin{figure}[ht]
\begin{center}
\resizebox{0.9\linewidth}{!}{%
   \includegraphics{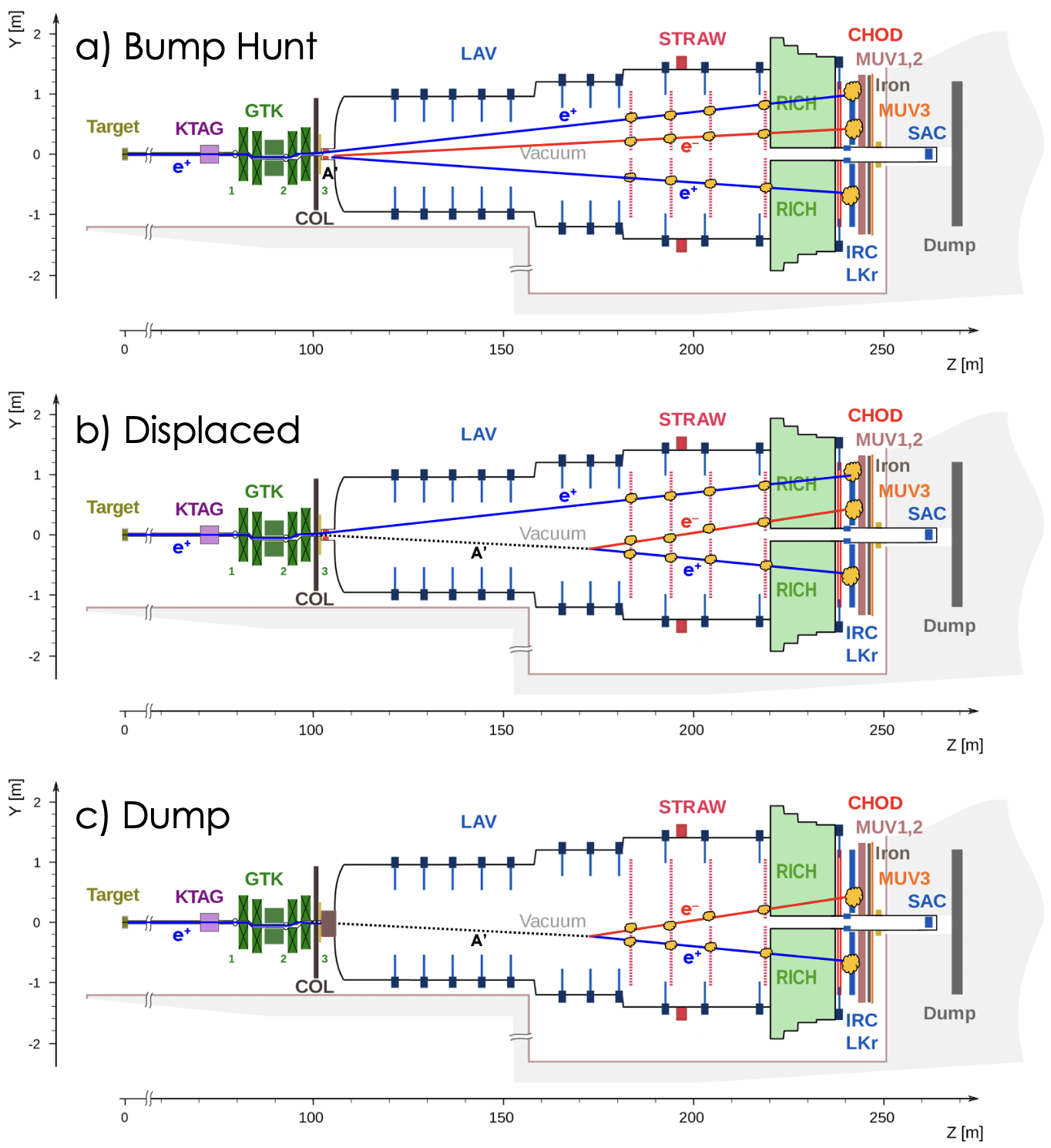}
}
\caption{NA62e$^+$'s visible decay search techniques. a) Bump hunt b) Displaced vertex c) Dump mode.} 
\label{fig:VisTech}    
\end{center}
\end{figure}

Fig.~\ref{fig:VisTech} illustrates the various techniques for detecting visible final states at NA62e+. In panels a) and b), the final state is obtained by considering only $A'$ bremsstrahlung production. Similar searches, originated by associated production where a photon replaces beam positron as final state particle, can also enhance the sensitivity. 
In dump mode, panel c), the accompanying particle is irrelevant as it is absorbed by the dump itself.

\subsection{Thin target: bump-hunt}

The bump-hunting technique, illustrated in panel a) of Fig.~\ref{fig:VisTech}, is employed to search for resonances within a smooth background. In this approach, a dark-sector particle produced at the target promptly decays into Standard Model particles. The reconstructed invariant mass is expected to exhibit a distinct peak superimposed on the smooth background distribution.
Several final states are available in a positron-on-target experiment depending on the production mechanism and the exotic particle decay of interest. In the case of NA62e+, we assume that the dominant decays are to lepton pairs, as hadronic decays are sub-dominant in this mass range. 
Limiting thus our analysis to leptonic decay products, the final state generated by both $A^{\prime}$ and ALPs with dominant electron couplings are:

\begin{enumerate}
    \item [a)] $e^+e^-\to X\to e^+e^-$ or $e^+e^-\to X \to  \mu^+\mu^-$ 
    \item [b)] $e^+e^-\to \gamma X \to \gamma e^+e^-$ or $e^+e^-\to \gamma X \to \gamma \mu^+\mu^-$ 
    \item [c)] $e^+N \to e^+N X \to N e^+ e^+e^-$ or $e^+N \to e^+N X \to N e^+ \mu^+\mu^-$ \end{enumerate}
In all the cases mentioned, a pair of leptons will be reconstructed by the NA62e+ spectrometer, enabling the computation of their decay vertex and invariant mass. 
In case the dominant coupling of an ALP $a$ is to photons, 
one has  the following final states: 
\begin{enumerate} 
    \item[d)] $e^+e^-\to a\to \gamma \gamma$ 
    \item[e)] $e^+e^- \to \gamma a \to \gamma \gamma \gamma$ 
    \item[f)] $e^+e^- \to e^+e^- a \to e^+e^- \gamma \gamma$ 
    \item[g)] $e^+N \to e^+N a \to N e^+ \gamma\gamma$ 
\end{enumerate}
The acceptance has been estimated with CalcHEP to be approximately 10\% using a Monte Carlo simulation of the process $e^+ e^- \to \gamma A’$ with the decay $A’ \to e^+ e^-$, requiring all final-state particles to fall within the calorimeter and the spectrometer acceptance. A similar acceptance is expected for the other decay channels, ensuring that NA62e+ maintains good detection capability across all final states.

The main source of background for visible $A'$ searches is the Bhabha scattering process, whereas for ALPs, it is the $e^+e^- \to \gamma\gamma$ process. At 75 GeV, the total Bhabha cross section within the NA62e+ angular acceptance is $\sim 9 \times 10^6$ pb, while the cross section for $ e^+e^- \to \gamma\gamma $ is $1.4 \times 10^6$ pb, posing significant challenges to visible decay searches.

Evaluating the reach for all visible final states necessitates a reliable assessment of the SM background. This will be addressed in future work, potentially utilizing the existing NA62 experiment Monte Carlo simulation to provide robust estimates.  

\subsection{Long-lived particles searches: dump mode}

Long-lived particles are natural dark sectors candidates.  Their weak couplings to SM particles inherently leads to extended lifetimes.
NA62 is uniquely equipped to directly measure the decay vertex positions of long-lived particles. The experiment's $70$-meter long decay region, coupled with the high Lorentz boost provided by the $75$ GeV positron beam, creates an ideal environment for such searches.
Two experimental strategies, illustrated in Fig.~\ref{fig:VisTech}, can be employed to distinguish long-lived particles from SM background. The displaced vertex technique, panel b), involves measuring the decay vertex and focusing the search on events exhibiting significantly displaced vertices. This approach, excluding mis-measured SM events, offers a background-free environment for sufficiently long lifetimes.

In situations where precise vertex reconstruction is challenging or unavailable, a high-Z material block can be employed as an absorber to eliminate SM background, panel c). Only long-lived particles that decay after traversing the absorber, will produce a detectable signal, effectively suppressing all SM background. 

\subsubsection{Thin target: displaced vertex}

NA62's multi track vertex reconstruction capabilities have been demonstrated numerous times in kaon decay analyses, resulting in a decay vertex resolution of $\sigma_{VTX}\sim$ 0.5 m. 

The displaced vertex technique, illustrated in Fig.~\ref{fig:VisTech} b), involves detecting di-lepton or di photon decays with a vertex located several meters away from the target. After complete energy reconstruction, the only remaining background consists of bremsstrahlung photon conversions, which can only occur outside the NA62 decay region which is maintained in high vacuum.

This technique will allow exploring a variety of dark sector models. In this section, we focus on DPs and ALPs. As demonstrated in thin target invisible searches, a leading sensitivity in DP case translates into a leading sensitivity in the ALPs with electron coupling.
Models with ALPs featuring pure photon coupling are expected to demonstrate significantly higher sensitivity due to their extended lifetime, $\tau = 64\pi/(m_a^3 g_{a\gamma}^2)$, which enhances the experimental advantage of having a 70 m long decay region.

\begin{figure}[h]
\begin{center}
  \centering
  \resizebox{0.9\linewidth}{!}{%
    \includegraphics{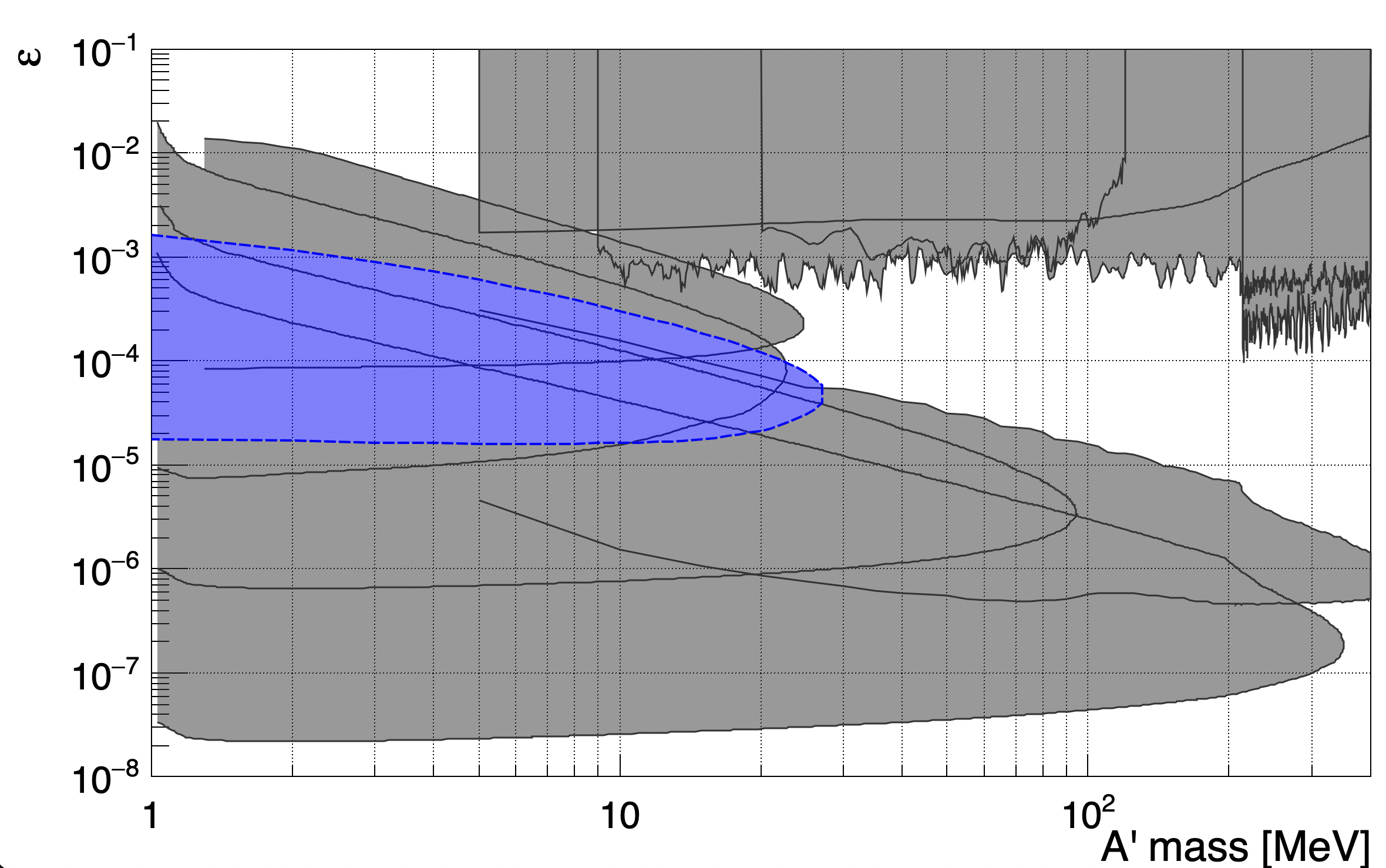}
  }
  \caption{NA62e+ $A'$ displaced vertex sensitivity (Blue) compared to existing limits shown in grey. Dump experiments NA64~\cite{NA64:2021aiq},
 E141~\cite{Riordan:1987aw}, KEK and Orsay~\cite{Konaka:1986cb,Davier:1989wz} KLOE~\cite{Anastasi:2015qla}, NA62~\cite{NA62:2023nhs}, bump hunt experiments KLOE~\cite{Anastasi:2015qla}, BaBar~\cite{BaBar:2017tiz}, NA48/2~\cite{NA482:2015wmo},LHCb~\cite{Aaij:2017rft}}

\label{fig:APDispVis}
\hfill
\end{center}
\end{figure}

In the DP  case we conservatively calculate the number of $A'$ produced 
in $A'$-strahlung only. 
Measuring the recoil positron momentum will allow additional constraints reducing background. 
After determining the decay length starting from the $A'$ width $\Gamma$, we estimate the probability of decay within the NA62 decay region as:
\begin{equation}
    P(L_{min},L_{max})=e^{-\frac{L_{min}}{L_{dec}}}-e^{-\frac{L_{max}}{L_{dec}}} 
\end{equation}

The value of $L_{min}$ depends on the NA62 vertex reconstruction capability and can be set at approximately 5$\times \sigma_{VTX}$. 
On the other hand, $L_{\text{max}}$ is constrained by the distance between the target and the first spectrometer station, which is approximately 70 m. 

With $2 \times 10^{14}$ 75 GeV e$^+$OT and a decay region defined by $L_{min}$ = 2.5 m and $L_{max}$ = 70 m, the parameter space accessible for $A'$ displaced vertex searches has already been explored by previous dump 
experiments (see Fig.~\ref{fig:APDispVis}). Competitive results for $A'$ displaced vertex searches in thin target mode require ~$2\times 10^{15}$ e$^+$OT, or an improved vertex position resolution. 

\subsubsection{Thick target: dump mode}

An alternative approach to leveraging the particle's lifetime is to use an absorber (dump) to suppress prompt Standard Model (SM) background. The NA62 experiment has demonstrated demonstrated successful operation in dump mode~\cite{NA62:2023nhs}.
Compared to proton on dump collected by NA62, the flux of positrons will be significantly lower, not exceeding approximately $10^{15}$ positrons on target. 
Nevertheless, the yield of dark sector particles is higher when positrons are used, due to $X$-strahlung. In addition, in proton dump mode the production is dominated by mesons decays reducing the momentum of dark sector particles and their decay length.

As demonstrated by the NA64 experiment, an absorber $\lesssim$ 50 cm  thick is sufficient to contain a positron beam of energy $\gtrsim$100~GeV. 
Dumping the positron beam at the beginning of the NA62 decay region would enable NA62e+ to achieve high acceptance, with a positron flux potentially up to three orders of magnitude higher than that of NA64\cite{NA64:2019auh,NA64:2021aiq}. 

\begin{figure}[ht]
\centering
  \centering
  \includegraphics[width=\linewidth]{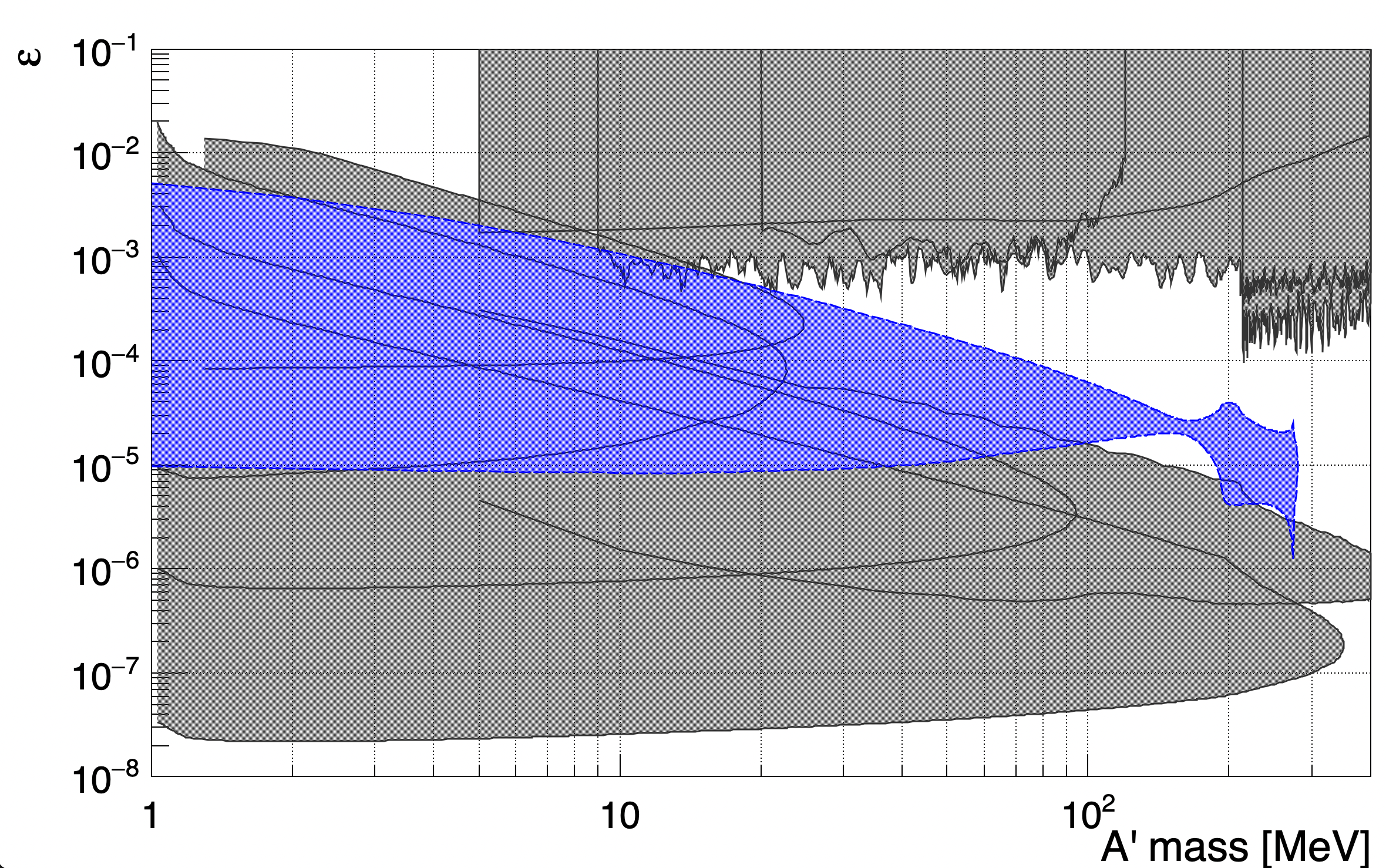}
  \caption{NA62e+ $A'$ sensitivity with $2\times10^{14}$ $e^+$OT and $d_{th}$=35\,cm (blue), compared to existing limits (gray). 
  }
  \label{fig:APdumpVis}
\end{figure}

In Fig.~\ref{fig:APdumpVis}, the blue filled area represents the expected NA62e+ sensitivity for $2 \times 10^{14}$ ~$e^+OT$ impinging on a tungsten dump with a thickness of $d_{\text{th}}$ = 35 cm (100 $X_0$). Decays of long lived meson downstream of the dump will be detected with extremely high efficiency by the NA62 apparatus.  
The gray filled region on the bottom indicates exclusions from beam dump experiments, while the region on the top corresponds to bump hunt exclusions form NA48/2 ($\pi^0 \to \gamma A^{\prime}$~\cite{NA482:2015wmo}),  BaBar ($e^+e^- \to \gamma A^{\prime}$~\cite{Lees:2014xha}), and KLOE~\cite{Anastasi:2015qla}.

The shape of the NA62e+ exclusion limit differs significantly from that of electron beam–dump experiments, as NA62e+ would be the first visible-dump experiment to use a positron beam.

\begin{figure}[ht]
  \centering
  \includegraphics[width=\linewidth]{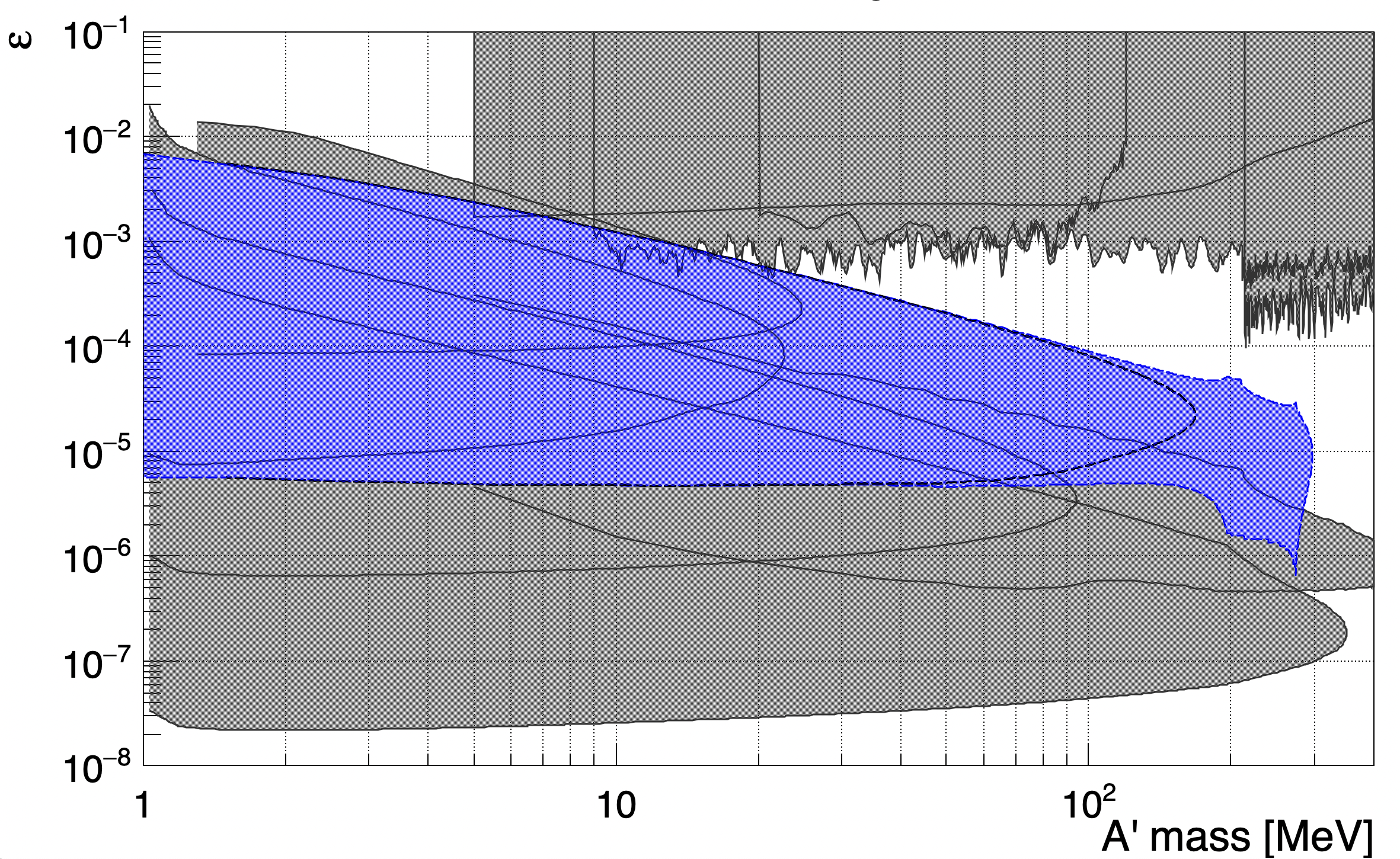}
  \caption{NA62e+ $A'$ sensitivity with $2\times10^{15}$ $e^+$OT and $d_{th}$=35\,cm (blue), compared to existing limits (gray).}
  \label{fig:APdumpVis_NoAcc}
\end{figure}

The smooth exclusion curves typically obtained with electron beams, shown in Fig. \ref{fig:APdumpVis_NoAcc} by the dashed black line, arise from considering $A'$-strahlung production only.
For the first time in the exclusions presented here the contributions from associated and resonant production (Fig.~\ref{fig:AProd}), along with the electron motion effects described in~\cite{Arias-Aragon:2024qji}, have been included in a visible-dump sensitivity calculation. These two new mechanisms extend the sensitivity reach into a region close to the CoM energy limit. Similar effects are present in the invisible channel sensitivity (see Fig. \ref{fig:InvMissE}), where they have already been documented in previous studies \cite{Marsicano:2018glj}.

Even with $2\times 10^{14}$ e$^+$OT, the visible dump mode allows access to a large unexplored region of the DP  parameter space. 

In Fig.~\ref{fig:APdumpVis_NoAcc}, the expected sensitivity for $2 \times 10^{15}$ e$^{+}$OT is illustrated. Under these conditions, the sensitivity is extended in the 100–200 MeV region, reaching the exclusion limit of the NA62 in dump mode~\cite{NA62:2023nhs}.

Raising the energy to 150 GeV is not a viable strategy, as this would lead to approximately a tenfold reduction in the positron rate, as shown in Fig.\ref{fig:PosProd}.

\begin{figure}[t]
    \centering
    \includegraphics[width=0.8\linewidth]{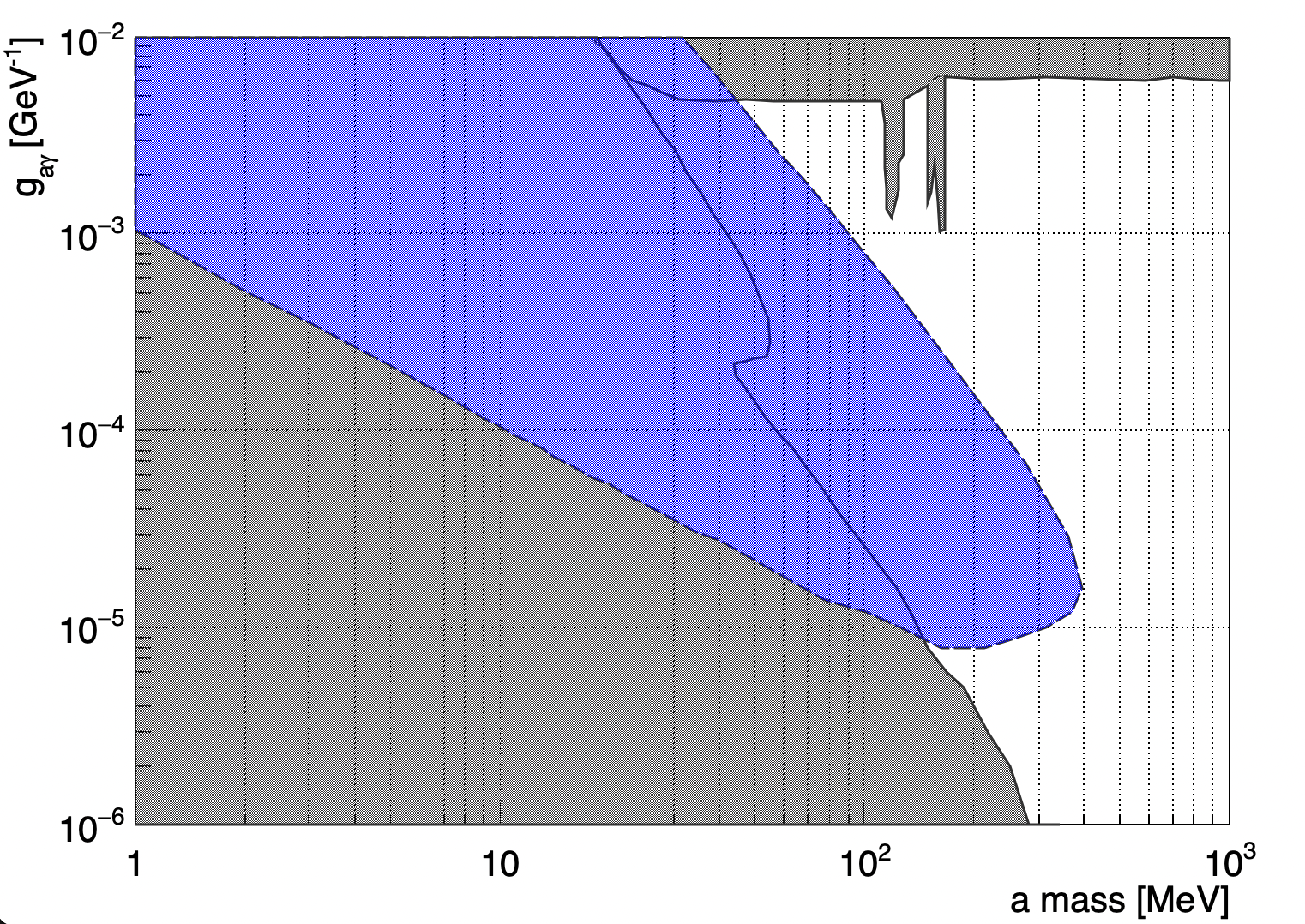}
    \caption{NA62e+ dump mode sensitivity to visible decays of ALPs with photon-dominant coupling (blue). In gray existing limits from LEP~\cite{OPAL:2002vhf}, NA64~\cite{NA64:PRL.125.081801}, Primex~\cite{Aloni:2019ruo}, and dump experiments as in Fig.\ref{fig:APDispVis}}
    \label{fig:ALPs_vis_Dump}
\end{figure}

Using Primakoff production rates obtained from \texttt{MadGraph5\_aMC}$@$\texttt{NLO}, with the requirement that the ALP carries more than 50\% of the beam energy, we derived the NA62e+ sensitivity in dump mode to ALPs with photon coupling, as shown in Fig.~\ref{fig:ALPs_vis_Dump}, assuming zero background. In this scenario, NA62e+ can achieve world-leading limits in the 50–400 MeV region. 
\section{Non-minimal dark sectors}

In addition to the minimal portal scenario already described, more elaborate models can produce observable final states in NA62e+. Detailed investigations of additional dark sector cases were performed in~\cite{PhysRevD.99.075001} for the LDMX experiment, including dark photons, Axion-Like Particles, dark scalars, milli-charged particles, and B-L new gauge boson models.

While expected limits from NA62e+ are not computed here, they can be extrapolated from the plots in~~\cite{PhysRevD.99.075001}. In general, limits from NA62e+ are expected to be stronger compared to electron beam ones for the same number of particles on target due to the additional contribution specific for positrons.
This section will address extensions with an additional dark Higgs, which is not covered in~~\cite{PhysRevD.99.075001}.

Non-minimal extension of the Standard Model (SM) by adding a new U(1)$_D$ gauge boson, $A'$, and a single complex scalar Higgs field, $\phi$, responsible for spontaneous symmetry breaking have been described in~\cite{Batell:2009yf}.
In the case were possible additional dark sector particles (such as a new dark matter candidate for instance) are heavy compared to the dark photon $A'$ and dark Higgs $h'$ mass scales, the phenomenology at accelerators is mainly driven by these two particles.

\begin{figure}[h]
 \begin{center}
 \resizebox{0.9\linewidth}{!}{%
   \includegraphics{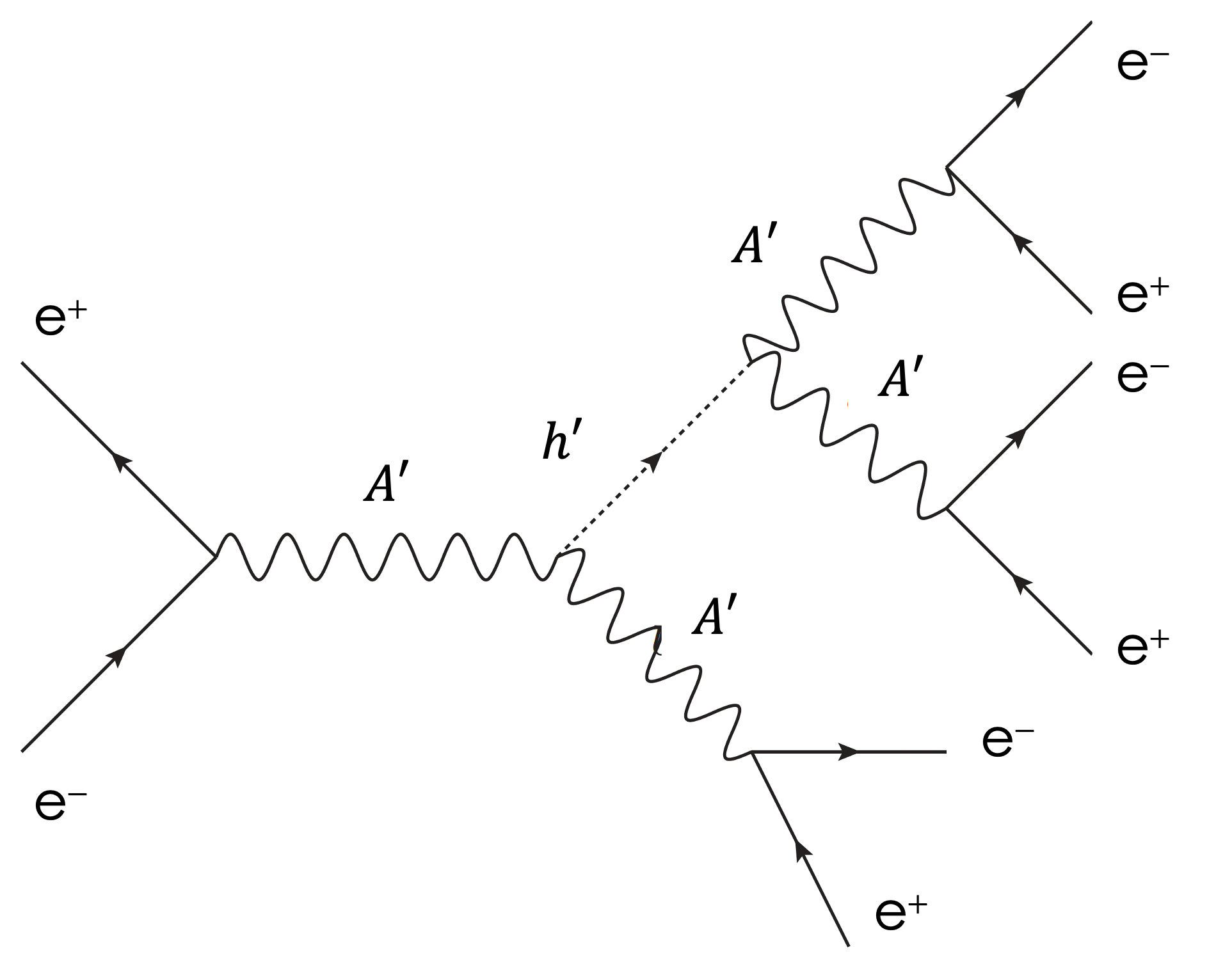}
 }
 \caption{The 6 leptons process $e^+e^-\to 3e^+e^-$ mediated by Dark Higgs.}
 \label{fig:DHto6L}    
 \end{center}
 \end{figure}

All interactions between the SM and the secluded sector occur through kinetic mixing of $U(1)_D$ with the Standard Model photon. 
The dark Higgs (DH) decay channels depend on whether it is heavier or lighter than the
dark photon $A'$. We will focus on
the case $m_{h'} > 2m_{A'}$, in which the $h'$ mainly decays to a pair of real dark photons. The decay is always prompt, since the lifetime only depends on the dark gauge coupling $g_D$ which is supposed to be of order one.
For c.m. energies below $\sim 200$ MeV, the phenomenology is simpler than that described in~\cite{Batell:2009yf}.  
Indeed, neglecting the possible decay to neutrinos, which is extremely suppressed at this energy, the only possible $A'$ decay is in $e^+e^-$ pairs.
The $h'$-strahlung process is one of the few tree level production mechanisms for $h'$. The amplitude is suppressed by only a single power of the kinetic mixing constant, making $h'$ production feasible to occur in the range of $\epsilon ~O(10^{-2} - 10^{-3})$. The complete decay chain of interest in NA62e+ is shown in Fig.~\ref{fig:DHto6L}.

The total cross section for the Higgs$'$-strahlung process, $e^+ e^- \rightarrow A' h'$, shown in Fig.~\ref{fig:DHto6L}, is~\cite{Batell:2009yf}:


\begin{equation}
\begin{aligned}
\sigma_{e^+ e^- \to A' h'}
&= \frac{\pi \alpha \alpha_D \epsilon^2}{3 s}
\left(1-\frac{m_{A'}^2}{s}\right)^{-2}
\sqrt{\lambda\!\left(1,\frac{m_{h'}^2}{s},\frac{m_{A'}^2}{s}\right)
}
\\
&\quad \Bigg[\lambda\!\left(1,\frac{m_{h'}^2}{s},\frac{m_{A'}^2}{s}\right)+ \frac{12 m_{A'}^2}{s}\Bigg] .
\end{aligned}
\end{equation}

\begin{figure}[h]
 \begin{center}
 \resizebox{1.\linewidth}{!}{%
   \includegraphics{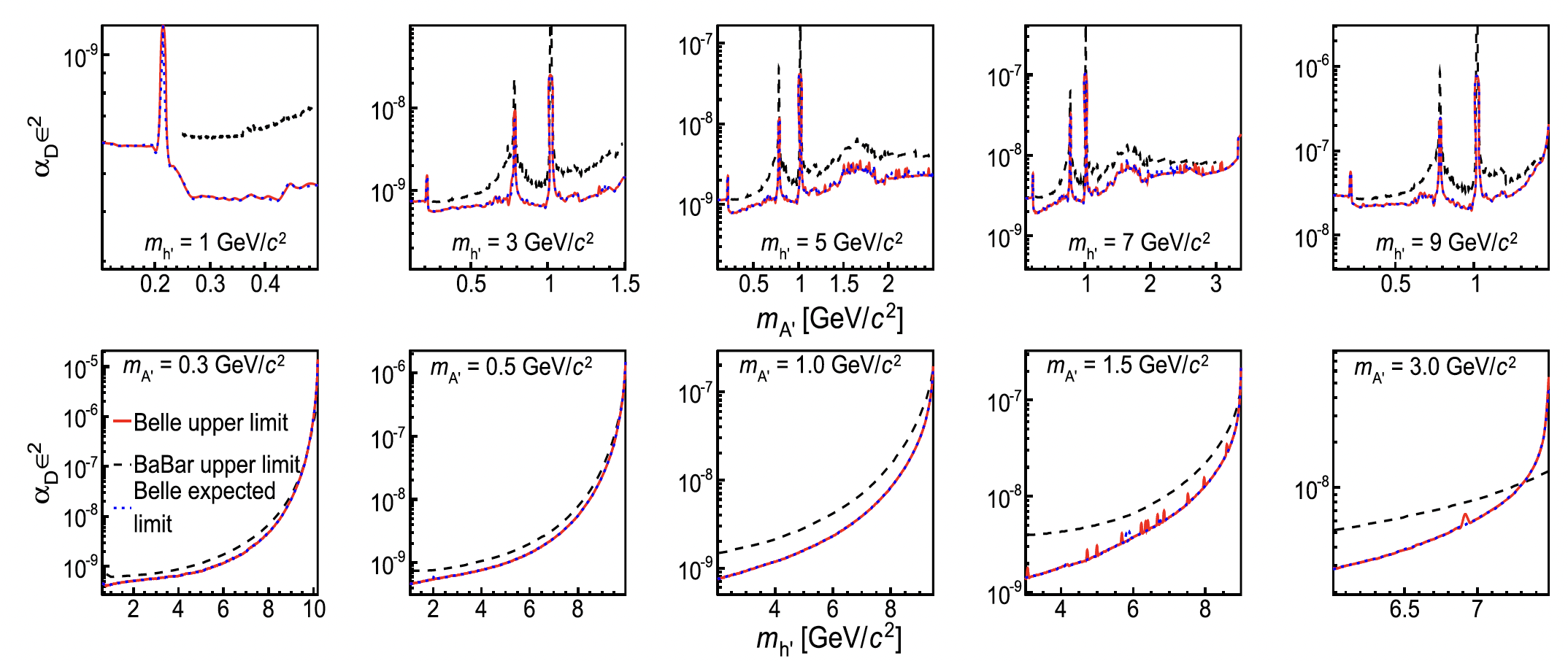}
 }
 \caption{90\% C.L. upper limit on the product $\alpha_D\epsilon^2$ versus dark photon mass (top row) and dark Higgs boson mass
(bottom row) for Belle (solid red curve) and BABAR (dashed black curve)~\cite{Jaegle:2015fme}.}
 \label{fig:DHLimits}    
 \end{center}
 \end{figure}

Using \( m_{A'} = 10 \) MeV, \( m_{h'} = 50 \) MeV, $\alpha_D=1 $, and $\varepsilon=1\times10^{-3}$ we obtain a total cross section of approximately 3.7 pbarn.  
The main Standard Model background process, \( e^+e^- \rightarrow 6e^\pm \), is significantly suppressed due to its proportionality to \( \alpha^6 \). A recent revision of the Equivalent Photon Approximation calculations has led to the following expression for the cross section~\cite{Ciafaloni:2020cic}:

\begin{equation}
\begin{aligned}
\sigma_{ee\to 6e}
&= \frac{\alpha^2}{6\pi^2}\,
\sigma_{(\gamma\gamma \to 2e^+e^-)}
\Bigg(\log^4\!\left(\frac{s}{m^2}\right)
+ \mathrm{A}\,\log^3\!\left(\frac{s}{m^2}\right)
\\
&\quad + \mathrm{B}\,\log^2\!\left(\frac{s}{m^2}\right)
+ \mathrm{C}\,\log\!\left(\frac{s}{m^2}\right) + \mathrm{D}\Bigg) .
\end{aligned}
\label{eqn:6l_FullEPA}
\end{equation}

where the constant coefficients have the following values:
\begin{equation}
A \sim -11.9 \quad
B \sim  22.62 \quad
C \sim  143.5 \quad
D \sim  -521.1
\end{equation} 
At 75 GeV positron beam energy the SM 6e background cross section value is 0.04 $\mu$barn.
Given the constraints of having each of the pair of lepton sharing the same invariant mass, namely the $A'$ mass, the background can be suppressed by several orders of magnitude as demonstrated by Babar~\cite{BaBar:2012bkw} and Belle~\cite{Jaegle:2015fme} studies.
We expect the DH to six leptons searches at NA62e+ to have no background. The strongest constraints on this models coming from Belle experiment~\cite{Jaegle:2015fme} are shown in Fig.~\ref{fig:DHLimits}.
The Belle search covered the region of 0.1$<m_{A'}/\text{GeV}<3.5$ and 0.2$<m_{h'}/\text{GeV}<$10.5, respectively. In the mass region $m_{h'}<$ 200 MeV NA62 should be able to provide the most stringent constraints.
In zero background hypothesis, the NA62e+ experiment could potentially achieve limits on $\epsilon \alpha_D$ as low as $\sim 10^{-8}$, profiting from the cross section enhancement due to the lower CoM energy compared to B-factories. However, more detailed studies on detector acceptance and actual sensitivity are beyond the scope of this paper. 

\section{Standard model physics}

Standard model cross sections for $e^+e^-$ collisions are per se an important part of the physics case. The NA62e+ luminosity at a c.m. energy below 300 MeV is unprecedented and can provide precise measurement of standard model observables in an almost unexplored regime.

\subsection{Mesons photoproduction and invisible mesons decays}

A recent study ~\cite{Schuster:2021mlr} has demonstrated that electron-on-target dark sector experiments are able to produce a large amount of light mesons through exclusive forward photoproduction. The mechanism requires a high energetic photon to be produced by SM bremsstrahlung in the interaction of the electron or positron with the experiment target. Subsequently, the photon undergoes an exclusive photoproduction process producing a single light vector meson (V).
Following the order of magnitude approximations in~\cite{Schuster:2021mlr}, the number of produced mesons ($N_V$) can be obtained as:
\begin{equation}
\begin{aligned}
N_V&= N_e f_{\mathrm{brem}} p_V , \hspace{1 cm}
p_V\approx \frac{9}{7}\,
\frac{\sigma_0^V X_0 f_{\mathrm{nucl}}^V}{m_p} ,\\
\sigma_{\gamma N \to V N}
&= f_{\mathrm{nucl}}^V\, A\, \sigma_0^V .
\end{aligned}
\end{equation}

where $\sigma^V_0$ is the cross section on a single nucleon $\sim 1\mu$b, and $f^V_{nucl}$ is an order 1 correction factor. 
If any of these mesons decay invisibly, it will produce a missing energy/momentum signal detectable by NA64 or LDMX~\cite{Schuster:2021mlr}. This allows us to translate missing energy/momentum limits into constraints on $BR(\rho,\omega,\phi,J/\psi)\to$Invisible.

\begin{table*}[htb]
\centering
\renewcommand{\arraystretch}{0.7}
\setlength{\tabcolsep}{2pt}
\begin{tabular}{l|ccc|ccc|cccc}
& $N_{e+OT}$ & $E_e$ & $f_{\text{brem}}$ & $\sigma_0^\rho$ ($\mu$b) & $\sigma_0^\omega$ ($\mu$b) & $\sigma_0^\phi$ ($\mu$b) & $N_\rho$ & $N_\omega$ & $N_\phi$ & $N_{J/\psi}$ \\ 
\hline
\rule{0pt}{1.1em}NA64\cite{NA64:2023wbi} & $1 \times 10^{12}$ & 100 & $0.5$ & $9$ & $0.8$ & $0.7$ & $2.5 \times 10^7$ & $2 \times 10^6$ & $2 \times 10^6$ & $2 \times 10^4$ \\
LDMX  I & $4 \times 10^{14}$ & 4 & $0.03$ & 23 & 5 & $0.4$ & $1.1 \times 10^9$ & $1.9 \times 10^8$ & $1.1\times10^7$ & -- \\
\hline
\rule{0pt}{1em}NA62e+ & $2 \times 10^{14}$ & 75 & $0.5$ & $9$ & $0.8$ & $0.7$ & $5 \times 10^9$ & $4 \times 10^8$ & $4 \times 10^8$ & $4 \times 10^6$ \\
\end{tabular}
\caption{Estimated meson yield from Tab. II in~\cite{Schuster:2021mlr}. We updated NA64 yield according to~\cite{NA64:2023wbi} and added an estimate for NA62e+ thick target mode.}
\label{tab:Mesonyields}
\end{table*}

In Tab.~\ref{tab:Mesonyields}, we updated the NA64 meson yields based on Ref.~\cite{NA64:2023wbi} and estimated those for NA62e+ in thick target mode by scaling the NA64 values (see Table II in~\cite{Schuster:2021mlr}).
The meson yields exceed $10^8$, except for $J/\Psi$. A null result in the missing energy experiment would translate into $\sim 10^{-8}$ limits on meson invisible decays, which is more than 1000 times stronger than current bounds (see Table I in Ref.~\cite{Schuster:2021mlr}).

In thin target mode, $\approx$15\% of the mesons yield in Tab.~\ref{tab:Mesonyields} is produced by the 500$\mu$m W target resulting in $\approx 4\times 10^7$ mesons/y, which will constitute one of the largest sample of $\rho$ and $\Omega$ meson ever obtained by photo production experiments.
In addition to the invisible decay limits, in thin target mode the energy of the photon that produced the meson can be derived measuring the momentum of the recoil positron. 
Tagging the mesons by their decay products and invariant mass, a measurement of the photoproduction cross section as a function of the photon energy can be derived with high   precision.
Future proton beam dumps such as SHiP will benefit a lot from this kind of measurements since, in their case, meson decays are an important source of DM. Typically, proton beam dump Monte Carlo simulations are tuned to match data but they just reach $\sim$25\% level agreement~\cite{Dobrich:2019dxc}.
Finally, by taking advantage of the NA62 detector apparatus, most of the decays of the $\rho$ and $\Omega$ mesons to leptons and pions, could be measured with higher precision than currently reported in the PDG review\cite{ParticleDataGroup:2024cfk}.

\subsection{The $e^+e^-\to\pi\pi$ cross section}

The $(g-2)_\mu$ anomaly is one of the most significant and longstanding discrepancies 
between an experimental measurement and the
Standard Model. 
Following the confirmation of earlier experimental results by E821~\cite{g-2_E821} obtained by the Muon g-2 Experiment  at Fermilab~\cite{g-2_G2M}, attention has increasingly turned to the theoretical predictions. These are carried out from first principles, by means of QCD lattice techniques, or via a data driven approach that uses as input the hadronic cross section. For this reason the measurement of the hadronic cross section at low energies is of great importance for the improvement of the theoretical error, and a particularly important role in this context is played by the dominant hadronic production process  $e^+e^- \to \pi^+\pi^-$. 

Currently, the most precise determinations of $e^+e^- \to \pi\pi$ as a function of $\sqrt{s}$ rely on the initial-state radiation technique, a method used e.g. by KLOE~\cite{KLOE-2:2017fda} and BaBar~\cite{BaBar:2009wpw, BaBar:2012bdw}.  Unfortunately, their results exhibit a $\sim3\sigma$ discrepancy.\footnote{The KLOE and BaBar determinations rely on the assumption that the scattering cross section for $e^+e^-\to \mu^-\mu^-$ is determined solely by SM processes, see~\cite{Darme:2021huc,Darme:2022yal} for a counterexample.} The hadronic cross section as a function of the c.m. energy can be also measured by varying directly the energy of the electron and positron beams, a  technique that has been used for the CMD-3 measurement. It is worrisome 
that the latest  CMD-3 result~\cite{CMD-3:2023alj} is not  consistent with either the one of KLOE nor BaBar.

Recently, Ref.~\cite{Arias-Aragon:2024gpm} proposed to leverage the momentum distribution of atomic electrons in positron annihilation on the electrons of a fixed target to scan over a large range of c.m. energies, while keeping the beam energy fixed. A proper high $Z$ material target can allow precise measurements of hadronic cross sections in a wide $\sqrt{s}$ region. It was shown in Ref.~\cite{Arias-Aragon:2024gpm} that with $10^{16}$ e$^+$OT a CERN north area experiment could yield a statistic larger than KLOE  up to c.m. energies of $\sim600\,$MeV. 

With a modest increase in the bending power of the achromats, NA62e+ could surpass the di-pion production threshold at a beam energy of 88 GeV. Utilizing a few $10^{14}$ positrons on target could achieve the statistical sensitivity required for a precise measurement of the di-pion cross-section near threshold, unlike previous measurements by KLOE and BaBar. By avoiding the reliance on ISR-based techniques, the NA62e+ measurement would be less affected by theoretical uncertainties, offering a robust and precise determination of this key cross-section. While detailed studies of this opportunity are outside the scope of this paper, we point out that the PID system of NA62 will allow for excellent lepton-hadron separation leading to very precise measurements at threshold. 

\subsection{Study of the di-muon production cross section}

A measurement of the $e^+e^- \to \mu^+\mu^-$ cross section with per-mille precision can be achieved during a one-year run in positron mode at NA62e+. The theoretical uncertainties for the $e^+e^- \to \mu^+\mu^-$ cross section are estimated at the 0.2\% level~\cite{kkmc}, demonstrating the significance of achieving such high precision. This measurement would provide valuable validation of QED in the low-energy regime, $\sqrt{s} \sim O(200)$ MeV, where no experimental data currently exists.  
In particular the observation of muon pair production with a beam energy below the threshold of 43.7 GeV for di-muon production off electrons-at-rest, will demonstrate unambiguously the effects of atomic electron motion and provide a  validation in high energy collisions of theoretical predictions~\cite{Arias-Aragon:2024gpm}\cite{Arias-Aragon:2025xcc}.

Finally, there is increasing interest in studying differential cross sections near the threshold, motivated by the proposed muon collider LEMMA scheme~\cite{Antonelli:2015nla}. In particular the beam emittance and the di-muon yield as function of the target material obtained in simulations need to be validated. The first tests with low statistical significance have been performed in the past using H4 positron beams~\cite{Amapane:2019oog}.  

\begin{figure}[h]
    \centering
    \includegraphics[width=0.45\columnwidth]{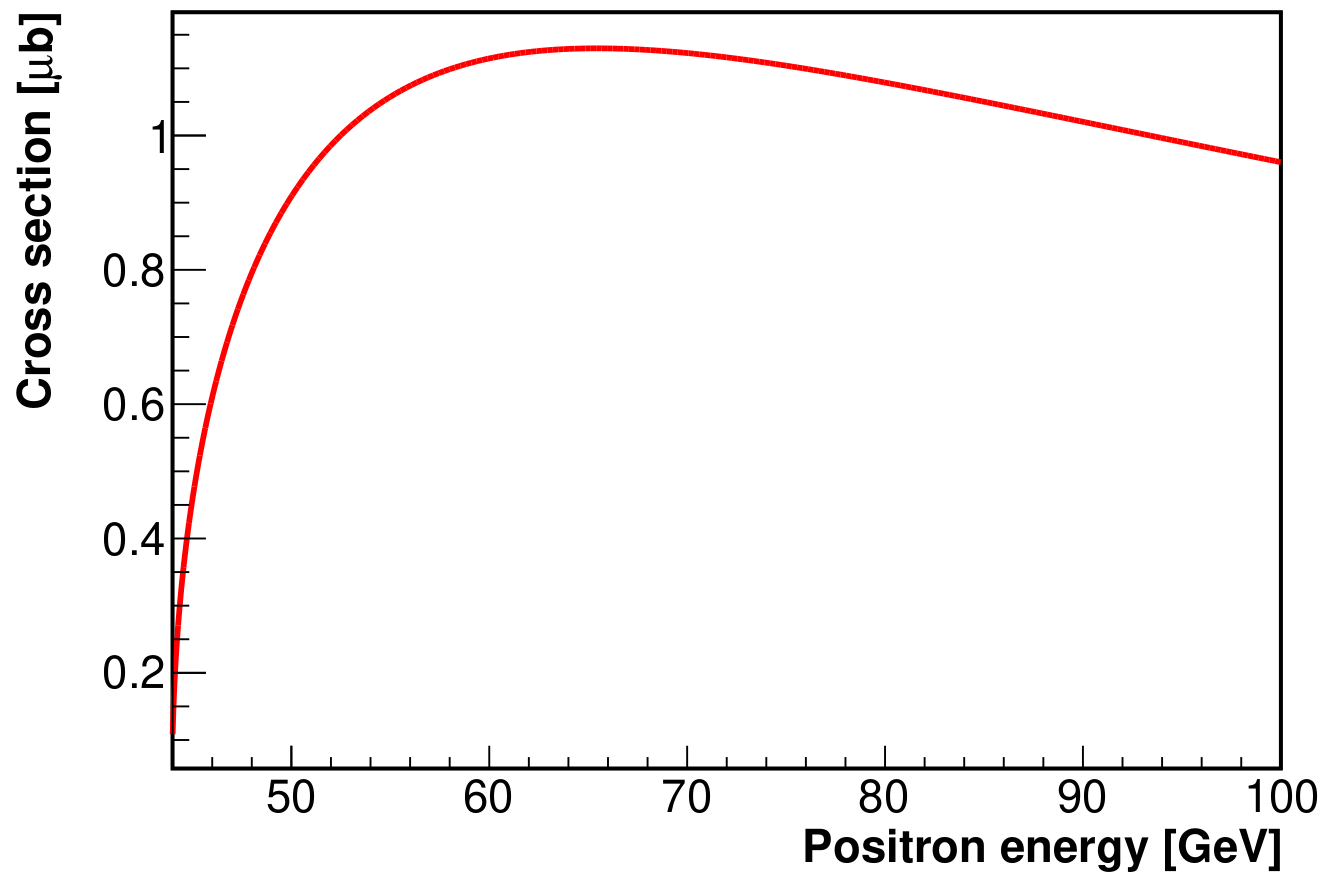}
     \includegraphics[width=0.45\columnwidth]{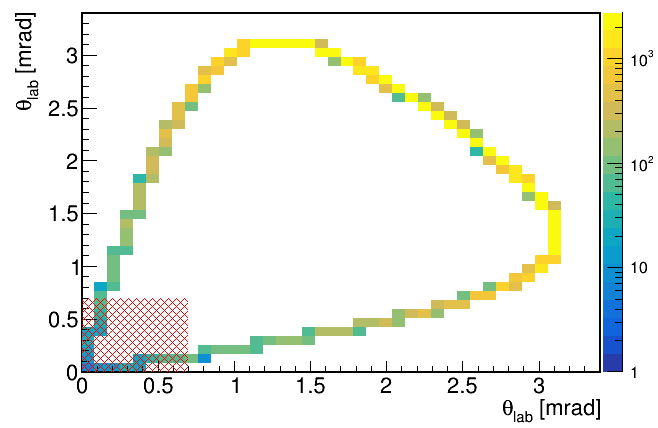}
    \caption{$e^+e^- \to \mu^+\mu^-$ annihilation cross-section at fixed target, as a function of the positron beam energy (left) and distribution of $\theta_{lab}$ for the two muons (right), with the straw tracker blind region in red .}
    \label{fig:sigma}
\end{figure}

The production cross section decreases at the production threshold of 45 GeV, see Fig.~\ref{fig:sigma}, therefore 75 GeV is an excellent compromise.\footnote{Notice that with enough statistics one could produce di-muons below the 45 GeV threshold by taking advantage of the tail of the atomic electron momentum distribution~\cite{Arias-Aragon:2024gpm}.}
The di-muon differential cross-section in the c.m. frame is equal to~\cite{pdg}:
\begin{equation}
\begin{aligned}
    \frac{d\sigma}{d\Omega} &= \frac{\alpha^2}{4s} \beta S(\beta) (2 - \beta^2 \sin^2\theta),\quad  S(\beta) = \frac{X(\beta)}{1-\exp(-X(\beta))} , \\ &\quad X(\beta) = \frac{\pi \alpha}{\beta} \sqrt{1 - \beta^2}
    \end{aligned}
\end{equation}
where $\beta = \sqrt{1 - \frac{4 m_\mu^2}{s}}$ and $S(\beta)$ is the Sommerfeld-Schwinger-Sakharov threshold enhancement factor from Coulomb rescattering~\cite{Brodsky:2009gx}.
The total cross-section is found by integrating the differential formula:
\begin{equation}
    \sigma = \frac{2 \pi \alpha^2 \beta S(\beta)}{s} \left( 1 - \frac{\beta^2}{3} \right)
\end{equation}
and its dependence on the positron beam energy is represented in Figure \ref{fig:sigma}.

At 75 GeV, the cross-section for the process $\sigma_{e^+e^- \to \mu^+\mu^-}$ is $1.1,\mu$b, which is very close to its maximum value. The maximum angle of the muons in the lab frame is approximately 3 mrad. Consequently, the acceptance is primarily limited by the lower bound of 0.73 mrad, determined by the inner hole (6 cm radius) of the straw tracker located 82 m downstream of the target.
The angular distribution $\theta_{lab}$ for the muon pair is shown in Figure \ref{fig:sigma}, along with the straw tracker’s blind region. 

The acceptance for $\theta_{lab} > 0.73$ mrad, calculated by integrating the differential cross-section is approximately 86\%. 
Assuming a $\sim$80\% selection and reconstruction efficiency, and including the $Z$ factor due to the target material, $3.02 \times 10^{-8}$ events/$e^+$OT are expected for the 0.5 mm Si target and 5 times more in the W case.

In one year of data collection with a 75 GeV beam and a positron flux of $2 \times 10^{14}$ $e^+$oT, approximately $N_{e^+ e^- \to \mu^+ \mu^-} \sim 6 \times 10^6$ events are expected, leading to a statistical uncertainty on the CS of $\sim$ 0.05\%. 
The background contribution from misidentified Bhabha scattering events, is strongly suppressed by the NA62 PID system, while contributions from bremsstrahlung photon conversion in the field of a nucleus
($\gamma N \to N \mu^+\mu^-$) and virtual photon bremsstrahlung from $e^+$ scattering off a $Z$ nucleus ($e^+ Z \to e^+ Z \gamma^* \to e^+ Z \mu^+\mu^-$) are suppressed by charged particle vetos and di-muon mass reconstruction.
With proper control of systematic uncertainties, a few per-mille level precision in the measurement of the cross-section is achievable in short time scale. 
\section{\label{sec:TMintro} True muonium observation at NA62e+}
Quantum electrodynamics (QED) predicts the existence of several bound states beyond standard atoms, including purely leptonic systems. The lightest such system, positronium ($e^+e^-$), was discovered decades ago and has been extensively studied.
True muonium (TM) ($\mu^+\mu^-$) and true tauonium ($\tau^+\tau^-$), however, have never been observed. This is primarily due to the lack of $e^+e^-$ colliders operating at the appropriate c.m. energy to exploit the enhanced resonant cross-section, as well as challenges related to dissociation effects in matter that complicate fixed-target production.
Atomic spectroscopy of true muonium is sensitive to contributions from hadronic vacuum polarization, such as those affecting the muon anomalous magnetic moment~\cite{tmhvp}, as well as to potential new physics involving lepton couplings~\cite{tmlamm}. Before pursuing TM spectroscopy, an initial observation of true muonium is necessary to enable preliminary studies, validate production and decay models, and lay the groundwork for future investigations.

Many pathways for its discovery are known, but one of the most promising is resonant production using a positron beam impinging on multiple thin lithium targets~\cite{h4}, which is also doable at NA62. A dedicated low energy data taking period at the di-muon threshold (43.7 GeV) could be performed as positron yield per proton on target is expected to be higher with respect to the nominal 75 GeV as shown in Fig.~\ref{fig:PosProd}. Together with possible true muonium discovery, the run will provide enhanced sensitivity to visible decays of low mass dark sectors candidates and useful information on the $e^+e^-\to \mu^+\mu^-$ cross section production at threshold.\\

True muonium has two spin states: para-TM (spin 0), which decays into $\gamma \gamma$, and ortho-TM (spin 1), which decays into $e^+e^-$. The ortho-TM state can be observed at NA62 by searching for its displaced decay vertices in $e^+e^-$.

The probability of producing the ground state ($1S$) in $e^+e^-$ interactions is $\epsilon_{1S} = 83\%$, and its lifetime at rest is 1.8 ps. This corresponds to a decay length of 11.3 cm for resonant production in a fixed-target setup~\cite{h4}. Higher excited states ($nS$) can also be produced, with probabilities $\epsilon_{nS}$ proportional to $n^{-3}$, which slightly increase the signal yield. The impact of these higher excited states, which also have higher ($\propto n^3$) lifetimes, and the potential for modifying the experimental setup to better exploit their production should be addressed in future studies, as these considerations are beyond the scope of this paper.

To compute true muonium production rate under realistic conditions, which involve non-negligible fluctuations in $\sqrt{s}$ and initial state radiation effects~\cite{h4}, the peak $e^+e^- \to TM$ cross-section $\sigma^{TM}_{peak}$ of 66.6 nb must be scaled down by considering the probability $p = 4.35 \times 10^{-4}$ that the beam c.m. energy falls within the energy window where bound states are allowed~\cite{Brodsky:2009gx}. This window has a width equal to the TM binding energy of 1.4 keV.
The effects of electron motion within the lithium target atoms and positron energy losses have also been included in the analysis~\cite{h4}. However, their estimated impact is negligible given the NA62e+ beam energy spread of approximately 1\%.

The most significant effect caused by matter is dissociation, as TM predominantly dissociates into $\mu^+$ and $\mu^-$ when passing through thick targets, with a dissociation cross-section of $\sigma_d \sim 13 Z^2$ b~\cite{lhcb}. For lithium, the dissociation length is $\mu_{d}^{-1} = 1.86$ mm.
A practical approach is to use an assembly of multiple lithium targets, each 4 mm thick ($\sim 2\mu_{d}^{-1}$), operated in vacuum. These targets should be spaced along the beam direction such that the majority of TM decays occur in the gaps between the targets.

The number of surviving TM atoms produced for one target per positron is~\cite{h4}:
\begin{equation}
        \frac{dTM}{de^+ dN_{\mathrm{target}}} =  \frac{p \, \sigma^{TM}_{peak}}{13 Z \, \mathrm{b}} (1 - e^{-\Delta z \mu_d})\epsilon_{nS} = 6.6 \times 10^{-13} \epsilon_{nS}
        \label{ntm_e-13}
\end{equation}

With a TM 1S decay length of 11.3 cm, a possible choice is to use 4 target cells, each featuring 4 beryllium foils spaced by 20 cm with tracking stations between them., i.e. 2 silicon detectors with a 20 cm spacing, for a total of 8 silicon detectors and 20 lithium foils (see Fig. \ref{fig:acc}). Including a 10 cm spacing between the last silicon detector of a cell and the first of the next cell, every cell is 110 cm long, for a total length of the target-trackers system of 4.4 m.\\

\subsection{Backgrounds discussion and detector requirements}

The $e^+e^- \to e^+e^-$ Bhabha scattering represents the main source of background, as follows by the comparison of cross-section and the fact that it has the same c.m. energy and final states of TM $e^+e^-$ decays~\cite{h4}. 
Note that background coming from di-muon production can be safely neglected due to the excellent particle identification capabilities of the NA62 experiment.
From the experimental point of view, the only differences between Bhabha scattering process and TM production and decay in $e^+e^-$ are the angular distribution in the c.m. frame and the non-zero decay length of TM.\\

\begin{figure}[h]
    \centering
      \includegraphics[width=0.45\linewidth]{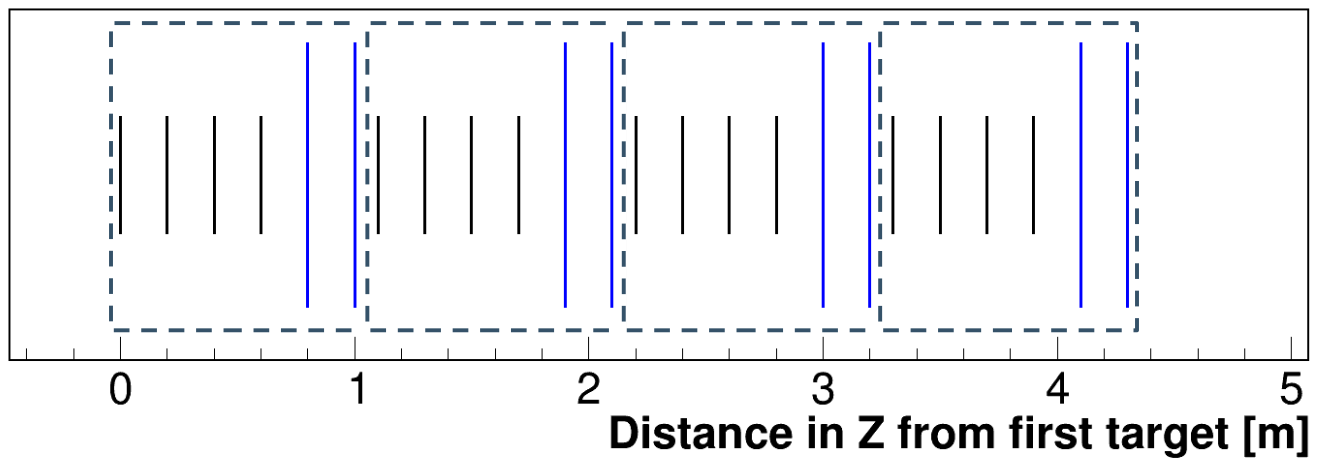}\hspace{0.4cm}
          \includegraphics[width=0.45\linewidth]{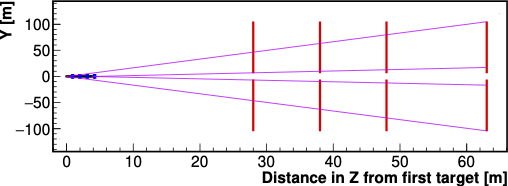}
    \caption{(left) Layout of the target-tracker setup: 4 cells with 4 lithium foils (black) and 2 silicon detectors (blue) each. (right) Layout of the targets together with the straw tube tracker modules (red)~\cite{NA62:2017rwk}. A distance of 28 m between the first target and the first tracker layer is chosen. Lines corresponding to minimum and maximum angles are drawn, respectively, in violet and green.}
    \label{fig:acc}
\end{figure}

An angular cut in the c.m. $\frac{\pi}{4} < \theta_{cm} < \frac{3}{4}\pi$ was chosen, resulting in a decrease of signal yield of a factor $\epsilon_{\theta_{cm}} = 62\%$ and a Bhabha cross-section of $\sigma_{Bh.} = 21 Z \mu b$. The minimum (maximum) angle in the lab frame of the $e^+/e^-$ originating from TM decays or Bhabha scattering are then 2.7 (16.6) mrad, corresponding to maximum (minimum) energies of 37.3 (6.4) GeV~\cite{h4}.\\
The NA62 straw tracker and LKr calorimeter are readily employable for this measurement. They are placed in a telescopic setup, meaning they have the same angular acceptance. The straw tracker has a hole at the center with a radius of 6 cm and an outer radius of 105 cm, with a distance between the first and last straw layer of 35 m~\cite{NA62:2017rwk}. 
By placing the first target at a distance of 28 m from the first straw layer, the acceptance for TM decay products with the above described $\theta_{cm}$ cut is nearly 100\%, as shown in figure \ref{fig:acc}. 
The only additional detectors required for the current NA62 setup are the silicon detectors inside the target. Per-layer resolution of $5 \mu$m and a material budget of 0.3\%$X_0$ are required, as estimated through simulations. Consequently, very thin monolithic pixel sensors, such as those planned for the ITS-3 ALICE upgrade or similar products~\cite{its3}, are necessary.

\subsection{Monte Carlo Simulations}
A simulation with $10^{14}$ $e^+$OT was conducted to demonstrate the feasibility of efficiently suppressing the QED background by distinguishing true muonium (TM) decays with displaced vertices from Bhabha scattering originating in the targets~\cite{h4}.

\begin{figure}[h]
    \centering
    \includegraphics[width=0.5\linewidth]{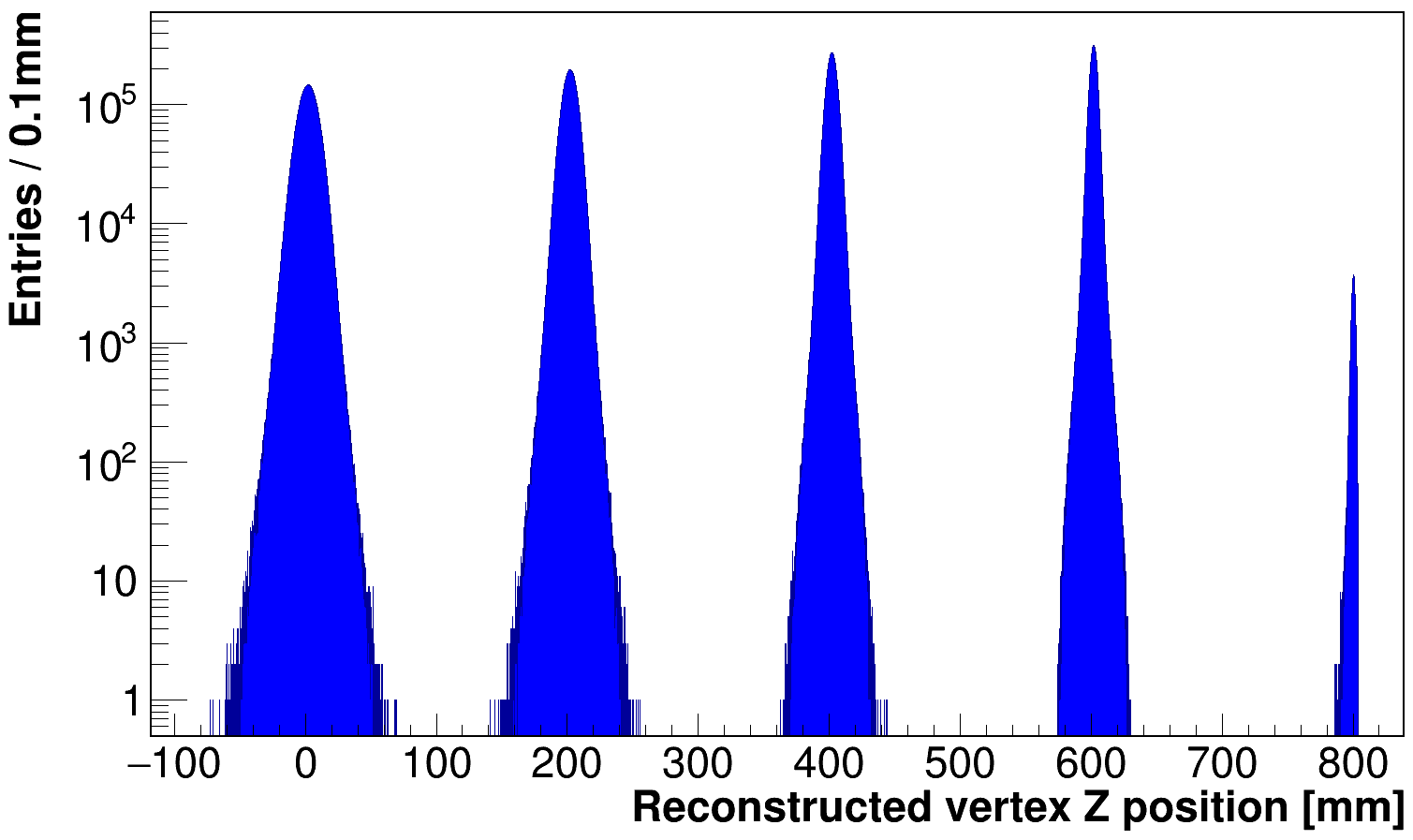}
    \caption{Reconstructed vertex $z$ position  after all other cuts for 4 targets of one simulated cell. The data at 800mm provide a very small yield and are due to fake vertices inside the first silicon detector.}
    \label{fig:z}
\end{figure}

A calorimeter resolution of $\sigma_E/E = 5\%/\sqrt{E[\mathrm{GeV}]} \oplus 10\%/E[\mathrm{GeV}] \oplus 1\%$, similar to that of the LKr calorimeter, was used. To identify Bhabha + TM events within the $\theta_{cm}$ acceptance, a series of pre-selection cuts were applied. The vertex $z$ of $e^+e^-$ candidates was then reconstructed, followed by additional quality cuts, resulting in a total reconstruction efficiency of $\epsilon_{reco} = 77.4\%$.
The identification of TM decays versus Bhabha scattering is achieved by applying cuts on the reconstructed vertex in the $z$-direction (see Fig.~\ref{fig:z}), after all other cuts. Background-free regions in $z$ are selected, so the expected background is determined by the simulated statistics, yielding $10^{-14}$ BG events per $e^+oT$ for one cell.

The cuts on the vertex $z$ position have different efficiency for TM events coming from different targets.
The average value for these efficiencies is 42.5\%~\cite{h4}.

\subsection{Discovery potential}
Depending on the experimental needs and data-taking time requirements for the primary goal of the experiment, which is dark sector searches, TM discovery could be conducted in either a short run (less than one month or even one week) using a dedicated target setup or a section of it, or in an approximately 3-month-long run using a less demanding setup with only one lithium target and two silicon tracker stations.

\subsection{Dedicated target setup: 16 target stations}

With the full target assembly discussed above, a discovery is possible in a very short time, making it the best choice in case of limited time at 43.6 GeV.

The global efficiency in equation \eqref{ntm_e-13} is evaluated by combining the probability to produce a 1S TM ($\epsilon_{1S} = 83\%$), the angular efficiency ($\epsilon_{\theta_{cm}} = 62\%$), the reconstruction efficiency ($\epsilon_{reco} = 77.4\%$), and the vertex-based selection efficiency ($\epsilon_v = 42\%$), reaching a value of $\epsilon_{tot} = 16.5\%$. After multiplying by the number of targets ($N_{target} = 16$), the value of selected TM per $e^+OT$ is evaluated to be:
\begin{equation}
\frac{N_{TM}}{e^+OT} = \epsilon_{tot} N_{target} \cdot 6.6 \times 10^{-13} = 1.74 \times 10^{-12}
\end{equation}\\
The expected background yield per $e^+oT$, given the simulated statistics of ($10^{14}$ $e^+$oT) per cell and on the number of cells $N_{cells}=4$ is:
\begin{equation}
 \frac{N_{BG}}{e^+OT}= \frac{N_{cells}}{10^{14}} = 4 \times 10^{-14}
\end{equation}
corresponding to a S/BG ratio of roughly 20.

In a data-taking period of 1 month out of 7 total months/year of data-taking, $e^+OT= 2\times 10^{14} / 7 = 2.8 \times 10^{13}$, therefore 48.8(1.1) signal (background) events are expected, corresponding to an 16.8$\sigma$ significance, which motivates hopes for a discovery also when including all experimental effects neglected in the toy simulation.
Moreover, a discovery level of significance of 6.7$\sigma$ could also be reached in a very short dedicated run of just one week.

\subsection{Parasitic setup with a single target station}

In order to reduce costs and the impact of the target installation on the main experiment schedule, a simplified target station can be used in parasitic mode.

In this case, only the last target of the assembly sketched above would be used, therefore $\epsilon_v = 56\%$ (see Table II of~\cite{h4}),
translating in $1.45 \times 10^{-13}$ TM/$e^+oT$ and $2.5\times10^{-15}$ BGK/$e^+OT$ events. 
In a period of 3 months out of 7 total months/year of SPS standard data-taking, 
$e^+OT= 2\times 10^{14} \times 2 / 7 = 5.7 \times 10^{13}$, therefore 8.3(0.14) signal (background) events are expected, corresponding to an 7.2$\sigma$ significance.

\section{Conclusions}

In this paper, we propose a physics program for the CERN NA, utilizing the existing NA62 detector and a newly developed high-intensity 75 GeV positron beam line. 
By capitalizing on the high proton flux expected in the post-LS3 era 
in the North Area, we suggest that positron fluxes $\sim2 \times 10^{14}\, e^+OT$/year can be achieved, using a small fraction of the protons delivered to the beam dump facility.

A dedicated feasibility and performance study by the CERN accelerator division on the development of a high-energy positron beam line in the NA would be highly 
desirable. Such a study would help determine the maximum achievable positron beam flux under various future North Area scenarios. 

The extended decay region, wide angular 
coverage of the veto systems, and the high-performance particle identification capabilities of the NA62 detector provide substantial advantages in suppressing background.

The proposed physics program focuses on dark sector searches using positrons on thin target collisions and positron dump mode. We have demonstrated that the NA62e+ experiment can achieve leading sensitivity to both visible and invisible decays 
of dark photons and ALPs within a short time frame.
The program can also include critical precision tests of the Standard Model, conducted concurrently with the thin target experiments. Precise measurements of hadronic cross-sections at low $\sqrt{s}$ can provide significant insights into the (g-2)$_\mu$ anomaly, contributing to a deeper understanding of this long-standing discrepancy.
Moreover, by reducing the c.m. energy to the di-muon production threshold ($E_{\text{Beam}} \sim 45 \, \text{GeV}$), $e^+e^- \to \mu^+ \mu^-$ interactions can be precisely studied, including the potential for the first observation of true muonium atoms.  

The NA62e+ physics program can be expanded to include additional dark sector models, such as inelastic dark matter, millicharged particles, and $B-L$ scenarios. This physics program can be accomplished together with the main BDF program, providing complementary results both in terms of parameter space and models explored with respect to the SHiP experiment\cite{SHiP:2015vad}. With the SHiP and NA62e+ program running simultaneously CERN NA would gain a world leading position in searching for feebly interacting particles, being able to access hadronic and leptonic production mechanisms, extended coupling and mass regions, and covering both lepto-phobic and lepto-phillic models at the same time.  
 
Finally, building a multi-technique positron-based dark sector experiment like NA62e+ offers a unique opportunity to probe the background limits of nearly all dark sector search techniques using positrons on target, establishing a foundation for future investigations at the extracted beam lines of the FCC-ee accelerator.

\section*{Acknowledgments}

We sincerely thank Lau Gatignon and Johannes Bernhard for their valuable discussions on the CERN NA beam lines and their future developments.
The work of M.R. was partially supported by CERN scientific associate program.
F.A.A., G.G.d.C. and E.N. are supported in part by the INFN ``Iniziativa Specifica" Theoretical Astroparticle Physics (TAsP). F.A.A. received additional support from an INFN Cabibbo Fellowship, call 2022.
The work of E.N. is  supported  by the Estonian Research Council, grant PRG1884.
Partial support from the CoE grant TK202 “Foundations of the Universe”, from the CERN and ESA Science Consortium of Estonia, grants RVTT3 and RVTT7, and  
 from the COST (European Cooperation in Science and Technology) Action COSMIC WISPers CA21106 is also acknowledged.
Sofia University is supported in part by the European Union - NextGenerationEU, through the
National Recovery and Resilience Plan of the Republic of Bulgaria, project SUMMIT BG-RRP-2.004-0008-C01 
and VK acknowledges partial support by BNSF KP-06-COST/25 from 16.12.2024 based upon
work from COST Action COSMIC WISPers CA21106 
supported by COST (European Cooperation in Science and Technology).

\bibliography{bibliography.bib}
\bibliographystyle{apsrev4-2.bst}

\end{document}